\documentclass[prl,twocolumn,showpacs,amsmath,amssymb,floatfix]{revtex4-1}
\usepackage{mathpazo,times,graphicx}
\usepackage[colorlinks=true,citecolor=blue]{hyperref}
\usepackage{cleveref}

\def\reftitle#1{``#1''}
\def\be{\begin{equation}}
\def\ee{\end{equation}}
\def\bse{\begin{subequations}}
\def\ese{\end{subequations}}

\makeatletter
\def\overl@ss#1#2{\vcenter{\offinterlineskip\ialign{$\m@th#1\hfil##\hfil$\crcr#2\crcr<\crcr } }}
\def\gl{\mathrel{\mathpalette\overl@ss>}}

\def\diag{\mathop{\rm diag}\nolimits}

\def\Real{\mathbb{R}}
\def\Complex{\mathbb{C}}

\def\Im{\mathop{\rm Im}\nolimits}
\def\Res{\mathop{\rm Res}\nolimits}
\def\re{\mathrm{re}}
\def\im{\mathrm{im}}
\def\d{\mathrm{d}}
\def\e{\mathop{\rm e}\nolimits}
\def\@#1{{\mathbf{#1}}}
\def\_#1{{\mathsf{#1}}}
\let\~=\tilde
\let\==\bar
\let\^=\hat
\let\<=\langle
\let\>=\rangle

\makeatother

\def\paragraph#1{\par\vskip0.4\smallskipamount\textbf{#1}\relax}

\let\eref=\eqref
\allowdisplaybreaks

\def\beginblue{\begingroup}
\def\endblue{\endgroup}
\def\blue#1{\begingroup#1\endgroup}

\def\thetitle{Soliton trapping, transmission and wake in modulationally unstable media}
\def\theauthors{Gino Biondini$^{1,2,*}$, Sitai Li$^2$ and Dionyssios Mantzavinos$^3$}
\def\theaddress{$^1$ State University of New York at Buffalo, Department of Physics, Buffalo, NY 14260, USA\\
$^2$ State University of New York at Buffalo, Department of Mathematics, Buffalo, NY 14260, USA\\
$^3$ University of Kansas, Department of Mathematics, Lawrence, KS 66045, USA}

\begin{document}
\title{\thetitle}
\author{\theauthors}
\affiliation{\theaddress}
\date{\small\today}

\begin{abstract}
Interactions between solitons and the coherent oscillation structures generated by localized disturbances \blue{via} modulational instability are studied
within the framework of the focusing nonlinear Schr\"odinger equation.
Two \blue{main} interaction regimes are identified based on the relative value of the velocity of the incident soliton compared to the amplitude of the background:
soliton transmission and soliton trapping.
Specifically, when the incident soliton velocity exceeds a certain threshold,
the soliton passes through the coherent structure and emerges on the other side 
with its velocity unchanged.
Conversely, when the incident soliton velocity is below the threshold,
once the soliton enters the coherent structure, it remains confined there forever.
It is demonstrated that the soliton is not destroyed, but its velocity inside the coherent structure 
is different from its initial one.
Moreover, it is also shown that, depending on the location of the discrete eigenvalue associated to the soliton,
\blue{these phenomena can also be} accompanied by the generation of additional, localized propagating waves in the coherent structure,
akin to a soliton-generated wake.
\end{abstract}
\maketitle

\paragraph{Introduction.}
The dynamics of nonlinear media affected by modulational instability
displays a number of interesting nonlinear phenomena
\cite{ZO2009}.
In particular, 
\beginblue modulational instability \endblue (i.e., the instability of the background with respect to long wavelength perturbations)
is the main mechanism for supercontinuum generation \cite{DudleyTaylor},
integrable turbulence \cite{zakharov2009,az2014,randoux2014},
and the formation of rogue waves \cite{onoratoosborne,solli,naturephys}.
\beginblue Modulational instability \endblue is described quantitatively by the one-dimen\-sional focusing nonlinear Schr\"odinger (NLS) equation, 
which is a universal model for the evolution of weakly nonlinear dispersive wave packets, and as such arises
in such diverse fields as water waves, plasmas, optics and Bose-Einstein condensates \cite{AS1981,Agrawal2007,IR2000,NMPZ1984,PS2003}.
Indeed, by linearizing the focusing NLS equation around the background, one can find
the range of unstable Fourier modes as well as their growth rate~\cite{benjaminfeir}.
But the linearization cannot capture the dynamics once the perturbations become comparable with the background,
which is referred to as the nonlinear stage of MI.

A qualitative explanation proposed for the nonlinear stage of \beginblue modulational instability \endblue is the formation of solitons
(in particular the so-called ``superregular breathers'' \cite{zakharovgelash,gelashzakharov,kibler}).
As shown in \cite{biondinifagerstrom}, however,
there exist broad classes of perturbations of the constant background that generate no solitons.
Therefore, 
solitons cannot be the main vehicle for the instability
(because all generic perturbations are linearly unstable, whereas for some perturbations no solitons are present in the solution).
Instead, in \cite{biondinifagerstrom} we showed that 
the signature of the instability lies in the nonlinear analogue of the unstable Fourier modes.
In \cite{biondinimantzavinos,CPAM2017} we then 
characterized the nonlinear stage of \beginblue modulational instability \endblue for localized {perturbations} of the constant background.
We showed that, generically, the nonlinear stage of \beginblue modulational instability \endblue 
displays universal behavior, with the $xt$-plane decomposing into two quiescent, or ``plane wave'' regions,
where the solution asymptotically equals the background up to a phase,
separated by a central region in which the leading order behavior is described by a slow modulation of the periodic, traveling wave
solutions of the focusing NLS equation. 
In \cite{biondinilimantzavinos} we further characterized the details of the asymptotic state
and showed that, for large times, the solution in the modulated oscillation region becomes a coherent collection of classical 
(i.e., sech-shaped) solitons of the NLS equation.
Moreover, in \cite{SIREV} we demonstrated that this kind of behavior is not limited to the NLS equation, but is instead 
a more general feature of nonlinear dynamical systems subject to \beginblue modulational instability\endblue.

A key requirement in \cite{biondinimantzavinos,CPAM2017,biondinilimantzavinos}, however, was that no solitons be present in the initial conditions.
A natural and important question is thus what happens when both solitons and a localized disturbance are simultaneously present.
The purpose of this work is to answer this question.
We do so by studying the interactions between solitons and the coherent oscillation structures generated by localized disturbances due to \beginblue modulational instability\endblue,
which for brevity hereafter we refer to as the ``wedge''.
We identify three interaction regimes based on the location of the discrete eigenvalue that gives rise to the soliton:
soliton transmission, soliton trapping, and a mixed regime \blue{in which the soliton transmission or trapping is} accompanied by 
the generation of additional, localized contributions in the wedge, which can be considered a soliton-generated wake.

\paragraph{NLS, solitons and \beginblue modulational instability\endblue.}
The starting point for our study is the focusing NLS equation, 
\vspace*{-1ex}
\be
iq_t + q_{xx} + 2(|q|^2 - q_o^2) q = 0\,,
\label{e:NLS}
\ee
where 
$q(x,t)$ 
is the complex envelope of a quasi-mono\-chromatic, weakly nonlinear dispersive wave packet,
and the physical meaning of the variables $x$ and $t$ depends on the physical context. 
(E.g., in optics, $t$ represents propagation distance while $x$ is a retarded time.) 
Here $q_o = |q_\pm|$ [with $q_\pm = \lim_{x\to\pm\infty}q(x,t)$]
is the amplitude of the nonzero background (NZBG).
The constant background solution of Eq.~\eqref{e:NLS} is simply $q_s(x,t) = q_o$.  
Equation~\eref{e:NLS} is the compatibility condition of the Lax pair \cite{ZS1972}
$\phi_x = X\phi$ and 
$\phi_t = T\phi$,
with 
$X = ik\sigma_3 + Q$ and 
$T = -i(2k^2+q_o^2-|q|^2-Q_x)\,\sigma_3 -2kQ$, where 
$\sigma_3 = \diag(1,-1)$ is the third Pauli matrix,  and
\vspace*{-0.6ex}
\be
Q(x,t) = \begin{pmatrix} 0 &q \\ - q^* & 0 \end{pmatrix}.
\vspace*{-1ex}
\ee
The first half of the Lax pair and $q(x,t)$ are referred to as the scattering problem and the potential, respectively.

The inverse scattering transform allows one to solve the initial-value problem for Eq.~\eqref{e:NLS} 
by associating to $q(x,t)$ time-independent scattering data via the solutions of the scattering problem.
Once the scattering data are obtained from the \beginblue initial condition\endblue,
$q(x,t)$ is reconstructed in terms of \blue{them} by inverting the scattering transform.
Specifically, the nonlinearization of the Fourier modes are the 
Jost eigenfunctions $\phi_\pm(x,t,k)$, which are the solutions of the Lax pair that reduce to plane waves as $x\to\pm\infty$.
In the case of NZBG, 
\vspace*{-0.6ex}
\beginblue
\be
\phi_\pm(x,t,k) = E_\pm(k)\e^{i\theta(x,t,k)\sigma_3} + o(1)\,,\quad x\to\pm\infty,
\label{e:jost}
\ee
\endblue
where
$E_\pm(k) = I + i/(k+\lambda)\,\sigma_3 Q_\pm$, and
\beginblue
\be
\theta(x,t,k) = \lambda x - 2k\lambda t\,,\quad
\lambda(k) = (k^2+q_o^2)^{1/2}\,.
\ee
\endblue
The Jost solutions are defined over the continuous spectrum $\Sigma = \Real\cup i[-q_o,q_o]$,
which is the image of the Fourier wavenumbers.
In particular, the range $i[-q_o,q_o]$ is the image of the modulationally unstable Fourier modes \cite{biondinifagerstrom}.
The scattering matrix $A(k)$ is defined by the scattering relation
\vspace*{-0.6ex}
\beginblue
\be
\phi_-(x,t,k) = \phi_+(x,t,k)A(k),\quad k\in\Sigma,
\ee
\endblue
and the reflection coefficient $r(k) = -a_{21}/a_{11}$ 
is the nonlinearization of the Fourier transform (see Appendix for details).

As usual, \blue{each} discrete eigenvalue, \blue{if} present, generates a soliton.
In the case of NZBG, the velocity of a soliton generated by a discrete eigenvalue at $k = k_o$ 
in the absence of localized disturbances is \cite{biondinikovacic}
$V_o = 2\sin\alpha\,(Z^2+1/Z^2)/(Z-1/Z)$, with $z(k_o)= iZ\,\e^{-i\alpha}$ and $z(k) = k + \lambda(k)$,
implying $Z>1$ and $\alpha\in(-\frac\pi2,\frac\pi2)$.

When no solitons are present, 
it was shown in \cite{biondinimantzavinos,CPAM2017} that an initial disturbance localized near $x=0$
generates a coherent oscillation structure confined to the wedge-shaped region $|x|< 4\sqrt 2 q_ot$, 
whereas outside this region the solution remains equal to the background value up to a phase.
\blue{The same approach as in \cite{CPAM2017} can be used to}
show that these features remain the same when solitons are present.

\begin{figure}[t!]
\smallskip
\centerline{\hglue-0.4em%
\includegraphics[height=0.245\textwidth]{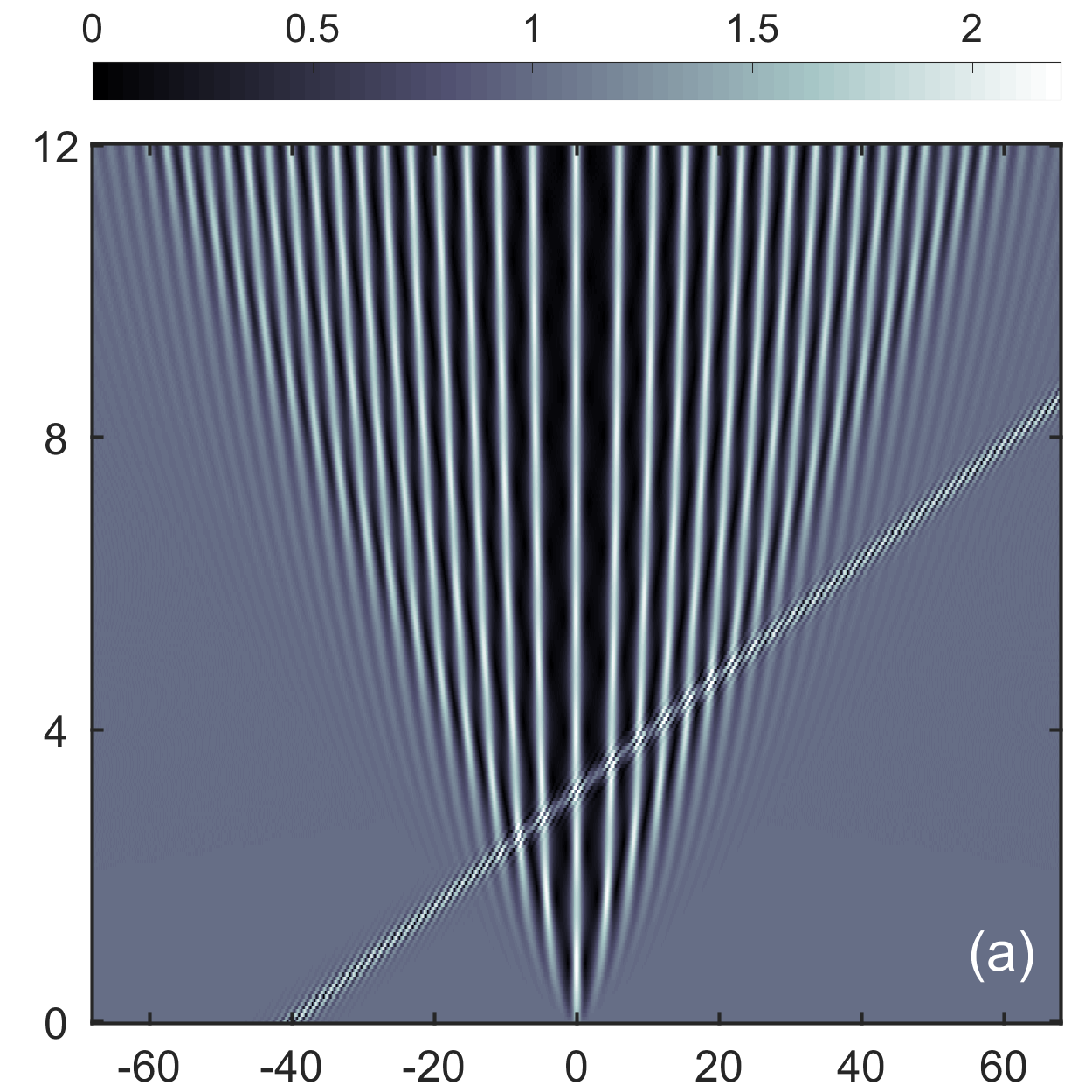}%
\includegraphics[height=0.245\textwidth]{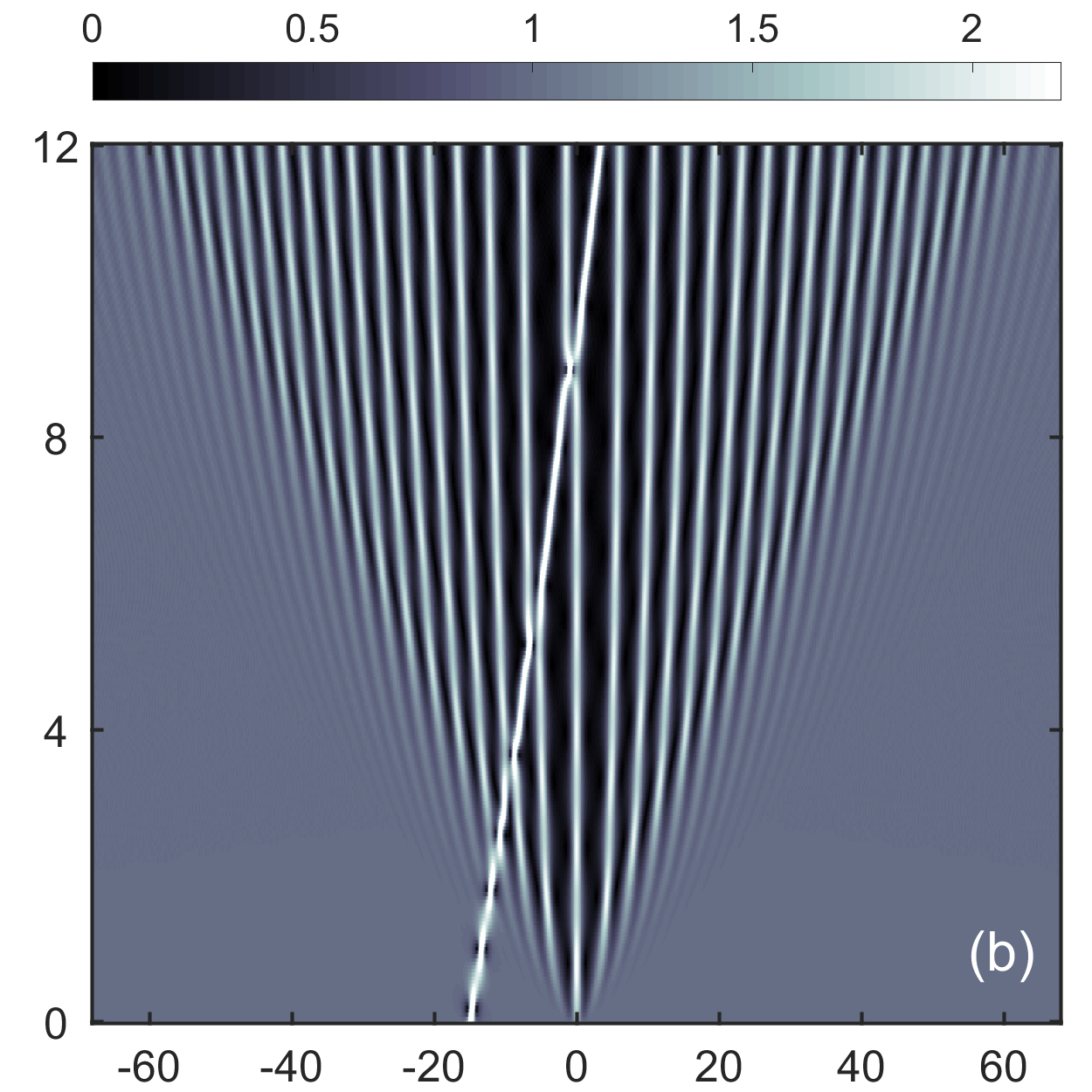}}
%
\caption{%
Density plots of numerical simulations of Eq.~\eref{e:NLS} with \beginblue initial conditions \endblue consisting of a soliton
plus a Gaussian perturbation of the constant background \blue{with $q_o=1$}.
The horizontal axis is~\blue{$x$}, the vertical axis is~\blue{$t$},
and the grayscale shows $|q(x,t)|$.
Recall that the wedge generated by an initial disturance localized at $x=0$ 
is confined to the region $|x|<4\sqrt2q_ot$ \cite{biondinilimantzavinos}.
\beginblue(a) \endblue $k_o = 3 + 0.5\,i$ (implying $V_o= 12.7$), resulting in a soliton transmission.
\beginblue(b) \endblue $k_o = 0.3 + 1.5\,i$ (implying $V_o= 1.63$), resulting in a soliton trapping.
}
\label{f:1}
\bigskip
\centerline{\hglue-0.4em%
\includegraphics[height=0.245\textwidth]{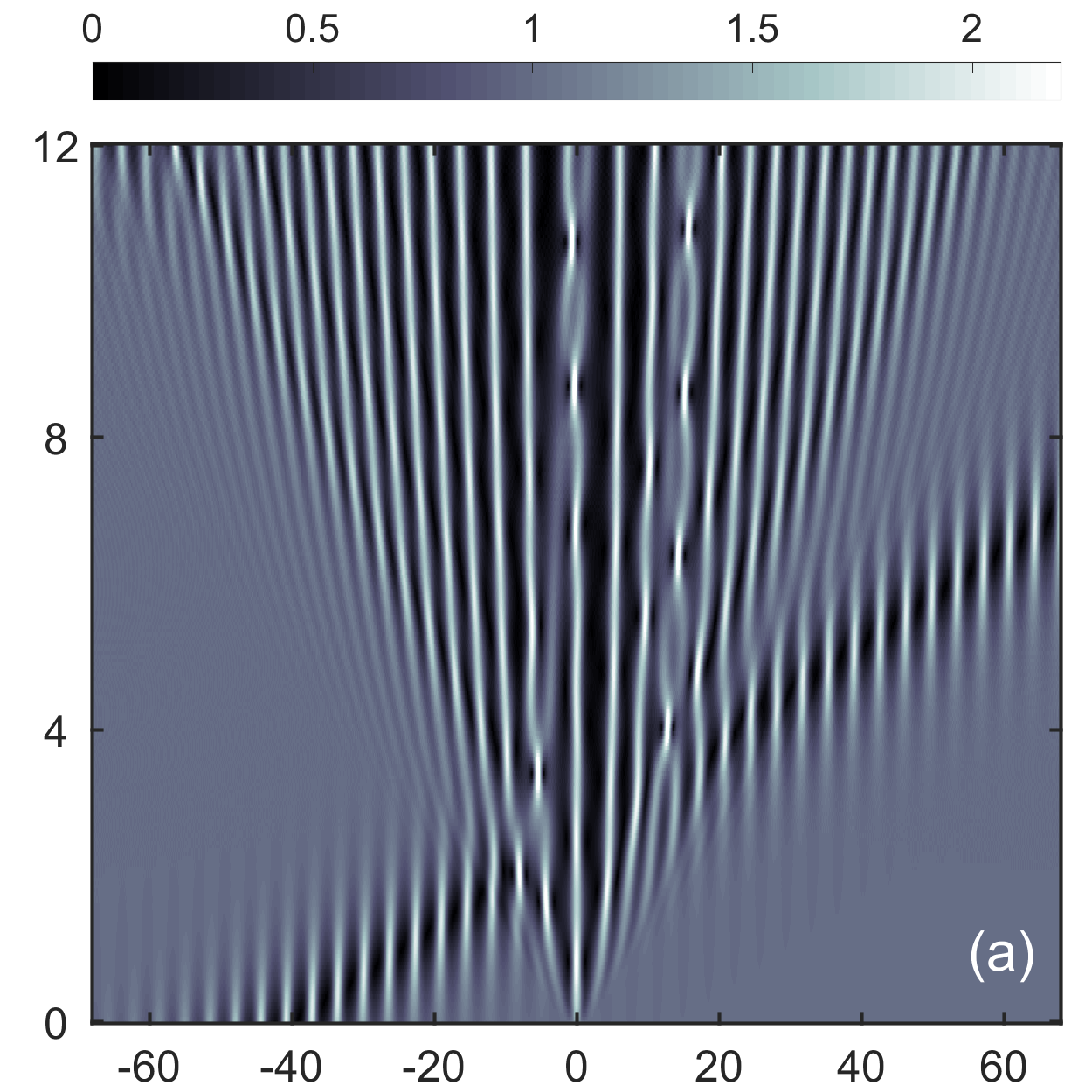}%
\lower1.2ex\hbox{\includegraphics[width=0.240\textwidth]{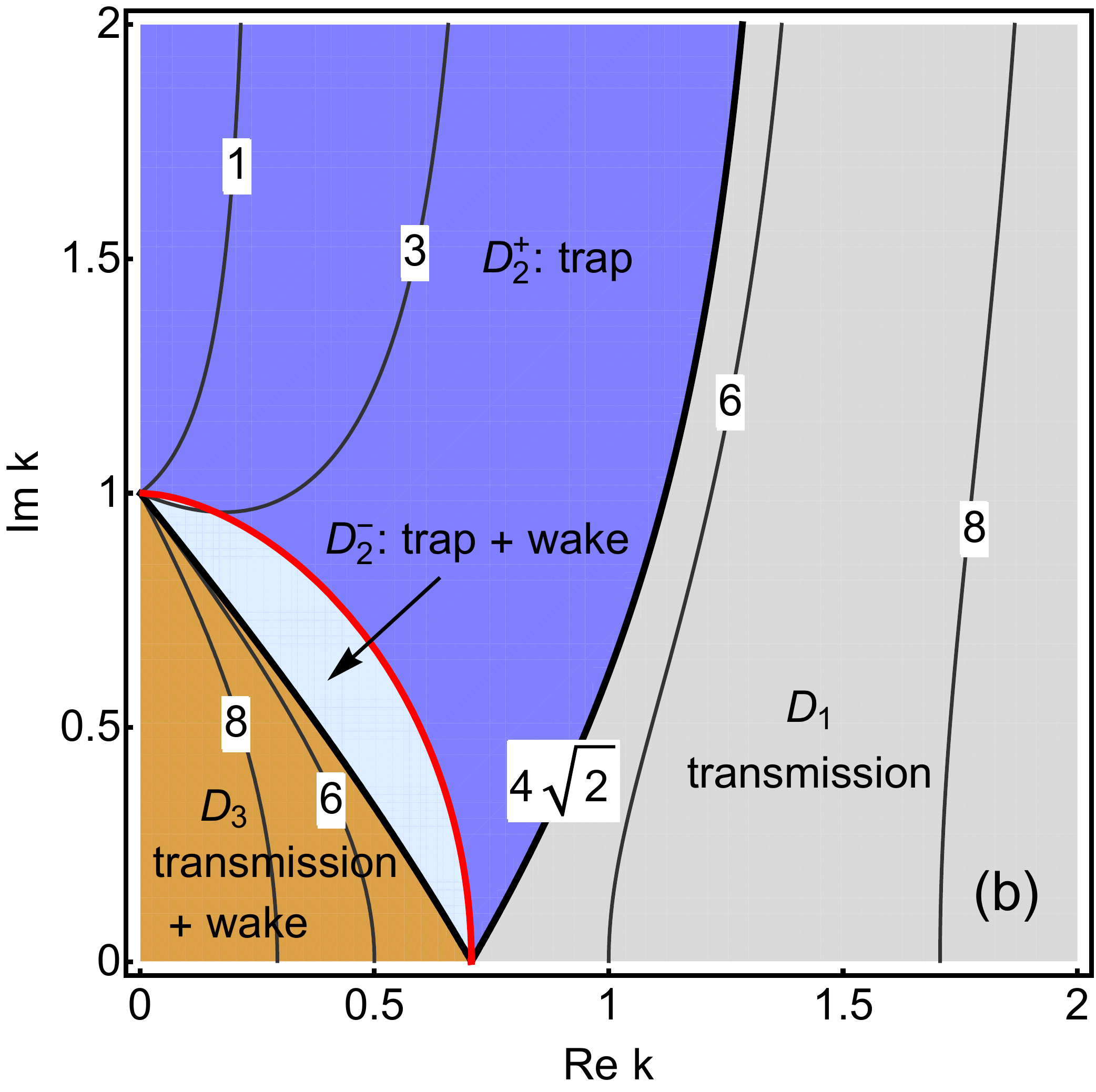}}}
\caption{%
\beginblue(a) \endblue Same as Fig.~\ref{f:1}, but for $k_o = 0.1 + 0.5\,i$ (implying $V_o= 15.5$), 
resulting in a mixed regime in a soliton transmission is accompanied by the formation of a wake in the wedge.
\beginblue(b) \endblue Contour lines of constant soliton velocity in the spectral plane 
and the domains $D_1$ (gray), $D_2$ \blue{(dark and light blue)} and $D_3$ (orange) of the spectral plane resulting in the \blue{various} outcomes.
\blue{(The red curve is determined by the modulation equations in \cite{biondinilimantzavinos}.)}
}
\label{f:2}
\kern-1.4\medskipamount
\end{figure}

\paragraph{Transmission, trapping and soliton wake.}
To study the interactions between solitons and localized disturbances, 
we performed careful numerical simulations of the focusing NLS equation~\eqref{e:NLS}
with a variety of \beginblue initial conditions \endblue consisting of a localized disturbance of the constant background initially placed at $x=0$
together with a soliton with initial velocity $V_o>0$ (generated by a corresponding discrete eigenvalue $k_o$)
and initially placed at a location $X_o<0$.
The kinds of possible outcomes are shown in Fig.~\ref{f:1} and Fig.~\ref{f:2}\beginblue(b) \endblue in three representative cases
corresponding to different choices for $k_o$.  
Those in Fig.~\ref{f:1} result in a soliton transmission and a soliton trapping, respectively,
whereas that in Fig.~\ref{f:2}\beginblue(a) \endblue results in a mixed outcome in which the soliton transmission is accompanied by 
the generation of additional contributions inside the wedge.
As we show next, 
all of these phenomena can be accurately described quantitatively by computing the
long-time asymptotics of solutions.

\paragraph{Solitons and long-time asymptotics.}
The \beginblue inverse scattering transform \endblue yields the solution of Eq.~\eqref{e:NLS} 
via the reconstruction formula \cite{biondinikovacic,CPAM2017}
\vspace*{-0.2ex}
\beginblue
\be
q(x,t) = -2i\lim_{k\to\infty}k M_{12}(x,t,k)\,,
\label{e:reconstruction}
\ee
\endblue
where $M(x,t,k)$ is the solution of a matrix Riemann-Hilbert problem \cite{Gakhov,TrogdonOlver}
defined in terms of the reflection coefficient, and, when discrete eigenvalues are present,
the corresponding poles and associated norming constants \cite{biondinikovacic,CPAM2017} (see Appendix for details).

As usual \cite{whitham,AS1981}, we compute the \beginblue long-time asymptotics \endblue along lines $x = \xi t$ with $\xi={}$const.
The difference between the present scenario and the one in \cite{biondinimantzavinos,CPAM2017} 
is the additional presence of poles in the \beginblue Riemann-Hilbert problem \endblue  coming from the discrete spectrum, 
i.e., the discrete eigenvalue at $k=k_o$ that produces the soliton.
\blue{Next we} briefly discuss how solitons arise \blue{in the calculation of the \beginblue long-time asymptotics\endblue.}

Both the jump condition and the residue conditions contain the phase $\theta(x,t,k) = \Theta(k,\xi)t$,
with $\Theta(k,\xi) = \lambda(\xi - 2k)$ \cite{biondinikovacic,CPAM2017} (see Appendix for details).
As in \cite{dkkz1996}, one can show that when the reflection coefficient is zero 
the poles give a \blue{vanishingly} small contribution 
to $q(x,t)$ \blue{as $t\to\infty$} for all $\xi$ such that $\Theta_\im(k_o,\xi)\ne0$.
Conversely, when $\xi$ is such that $\Theta_\im(k_o,\xi)=0$, the poles give an $O(1)$ contribution to the solution.
As a result, the soliton velocity is simply the value of $\xi$ such that $\Theta_\im(k_o,\xi)=0$.
When $q_o=0$ (i.e., with zero background), the phase reduces to $\Theta(k,\xi) = k\xi- 2k^2$.
In this case the above criterion recovers the familiar expression $V_o = 4k_\re$, with $k_o = k_\re + ik_\im$.
When $q_o\ne0$ instead (i.e., with NZBG), the same criterion yields
$V_o = 2(k_\re + k_\im\,\lambda_\re/\lambda_\im)$, with
$\lambda(k_o) = \lambda_\re + i\lambda_\im$,
which coincides with the expression given earlier.
Thus, one can identify the soliton velocity without computing the solution of the NLS equation.

Figure~\ref{f:2}\beginblue(b) \endblue shows \blue{the contour lines of the} soliton velocity in the spectral plane with NZBG.
Note that the curves of constant soliton velocity are simply \blue{given by} $\Theta_\im(k,\xi)=0$ for different values of $\xi$.
These curves touch the real $k$-axis twice for $V_o>4\sqrt2q_o$, once for $V_o = 4\sqrt2q_o$ and never for $V_o<4\sqrt2q_o$.
This feature affects the calculation of the \beginblue long-time asymptotics\endblue.
Also, Fig.~\ref{f:2}\beginblue(b) \endblue shows that the contour line $V_o = 4\sqrt2q_o$ divides the spectral plane into three domains:
$D_1$ (gray region), where $V_o>4\sqrt2q_o$ extending to infinity;
$D_2$ (\blue{dark and light blue regions, respectively $D_2^+$ and $D_2^-$}), where $V_o<4\sqrt2q_o$; 
and $D_3$ (orange region), where $V_o>4\sqrt2q_o$ touching the segment $[0,iq_o]$. 
Below we show that \blue{these regions correspond to} the locations of discrete eigenvalues that 
result in soliton transmission, trapping, and the mixed regimes.

Since $V_o>0$, 
the results of \cite{biondinimantzavinos,CPAM2017} also apply in our case for $x<0$, 
and the solution is unaffected by the presence of the soliton as $t\to\infty$,
apart possibly from an overall constant phase.
Thus, hereafter we restrict ourselves to computing the \beginblue long-time asymptotics \endblue for $x>0$.

\paragraph{Soliton transmission.}
Consider a discrete eigenvalue $k_o\in D_1$. 
Recall that the boundary of the wedge is given by the lines $x = \pm 4\sqrt2q_ot$, 
and the range $|\xi|>4\sqrt2q_o$ is the plane wave region.
The value $\xi = V_o$ lies in this range.
\blue{Even in the presence of a discrete eigenvalue, for $\xi\ne V_o$ the calculations proceed as in \cite{CPAM2017},
and the discrete eigenvalue yields no contribution to the solution in the \beginblue long-time asymptotics\endblue. 
Conversely, for $\xi = V_o$ it yields a leading-order contribution that results in the soliton, as before.
In other words,}
the \beginblue long-time asymptotics \endblue predicts that the soliton appears as a localized traveling object outside the wedge,
and that the soliton velocity after the interaction coincides with $V_o$,
in perfect agreement with the numerical results. 
Moreover, the \beginblue long-time asymptotics \endblue also recovers the asymptotic phase difference of the solution as $x\to\pm\infty$, consistently
with the constraint coming from the discrete eigenvalue \cite{biondinikovacic}.

\begin{figure}[b!]
\smallskip
\centerline{\hglue-0.4em%
\includegraphics[height=0.245\textwidth]{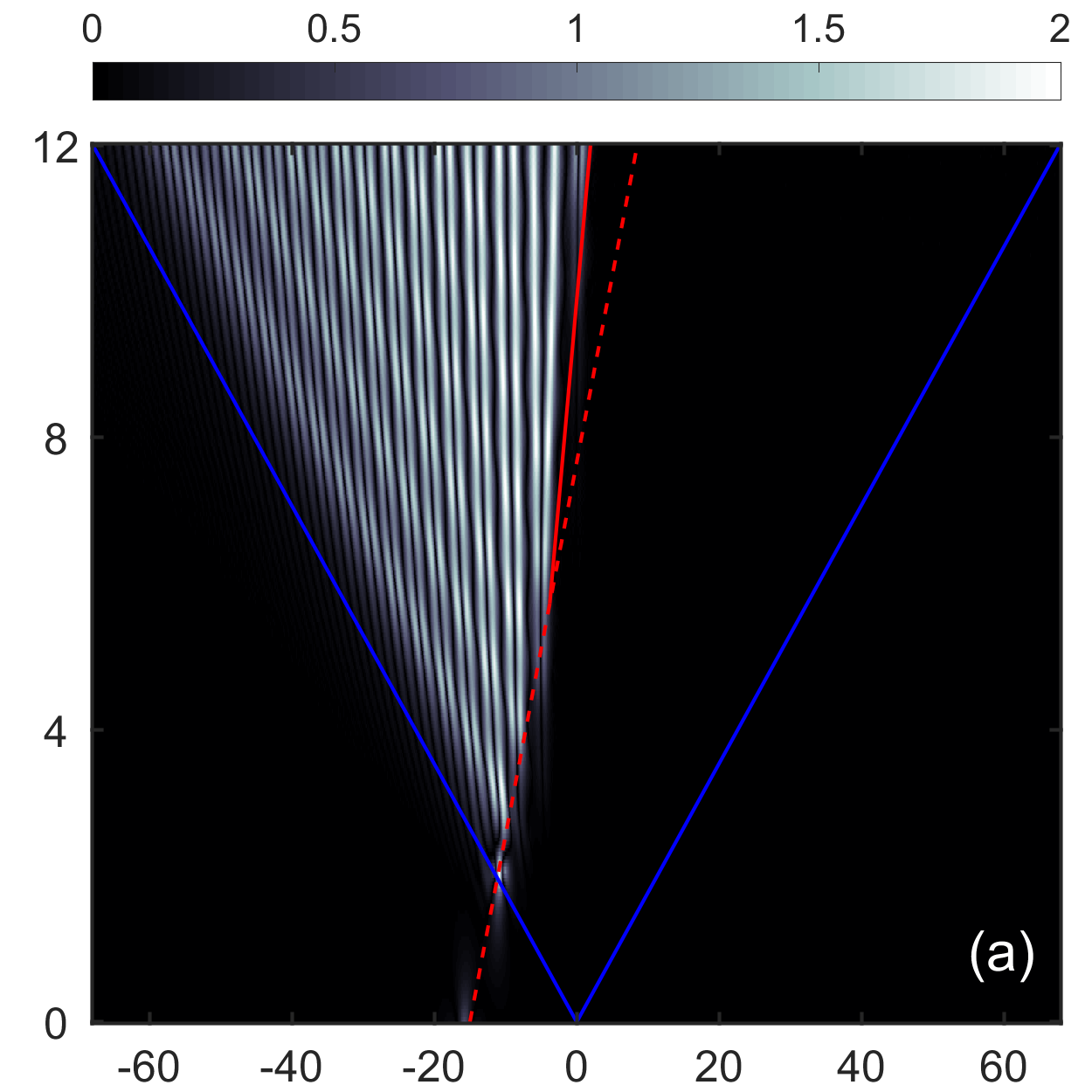}%
\includegraphics[height=0.245\textwidth]{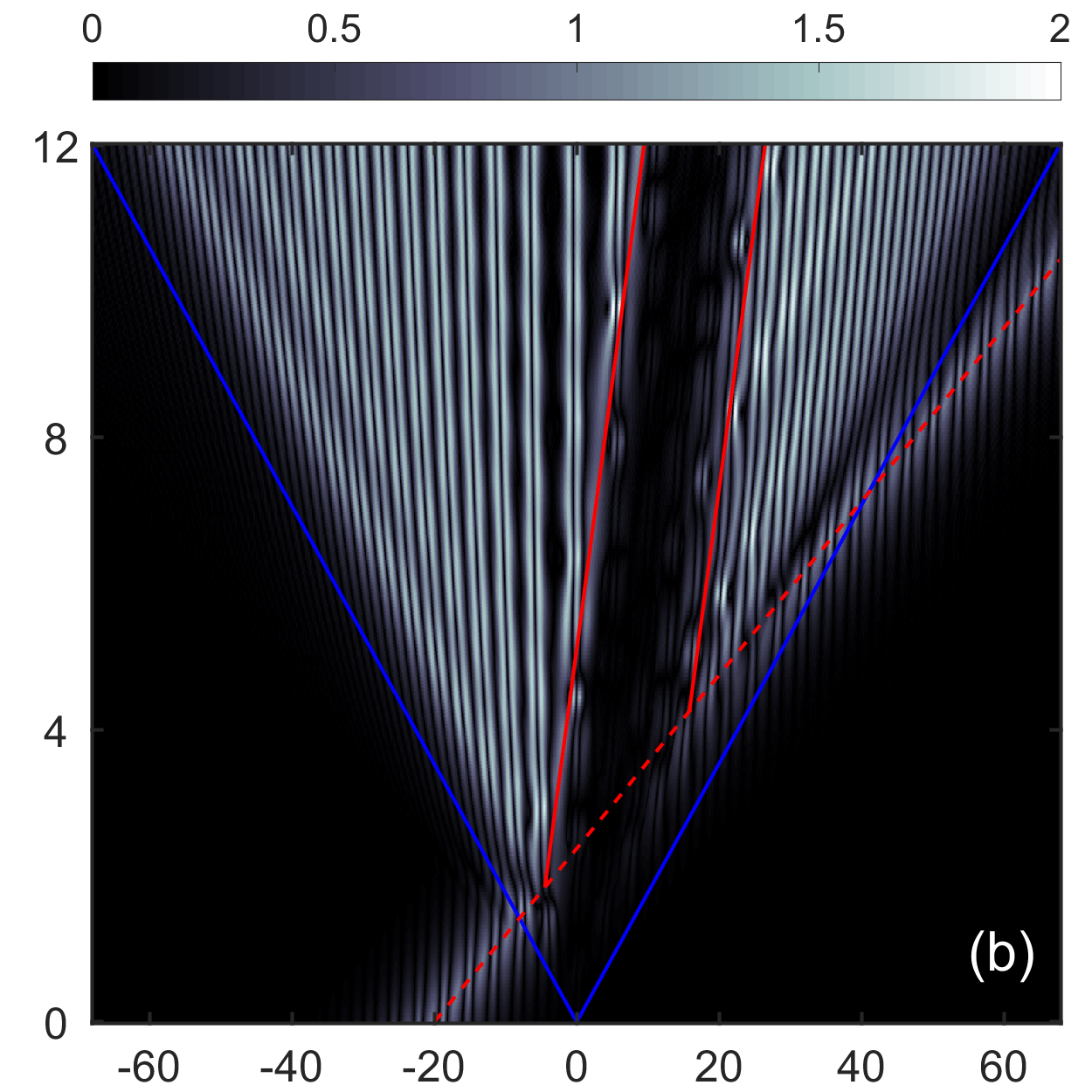}}
\caption{%
Density plots of the difference between the amplitude of a\break solution with a soliton and that of a solution without the soliton 
(i.e., generated by just the localized disturbance),
demonstrating the presence of the soliton inside the wedge, the change in the soliton velocity and the generation of the soliton wake.
\beginblue(a) \endblue $k_o = 0.1 + 1.02\,i\in D_2$ (implying $V_o = 1.95$), resulting in a soliton trapping.
\beginblue(b)~\endblue $k_o = 0.2+0.5\,i\in D_3$ (implying $V_o= 8.42$), 
resulting in a mixed regime comprising a soliton transmission and a soliton-generated wake.
Blue lines: the boundaries $x=\pm 4\sqrt2q_ot$ of the wedge.
Dashed red lines and solid red lines: trajectories corresponding to $V_o$ and to $V_*$, respectively.
}
\label{f:3}
\kern-\smallskipamount
\end{figure}

\paragraph{Soliton trapping and velocity change.}
The outcome changes when \blue{the soliton is generated by $k_o\in \smash{D_2^+}$}.
In this case $V_o<4\sqrt2q_o$, hence now the value $\xi = V_o$ occurs inside the wedge.
Thus, the \beginblue long-time asymptotics \endblue now predicts that \blue{no soliton is present} in the plane wave region,
consistently with the numerics.
A natural question is then whether the soliton is destroyed by the interaction or whether it persists inside the wedge.
To this end, 
Fig.~\ref{f:3}\beginblue(a) \endblue shows the difference between the amplitude of a solution with a soliton present
and that of a solution in which the soliton is absent.
The permanent change across the wedge in Fig.~\ref{f:3}\beginblue(a) \endblue clearly demonstrates
the persistent presence of the soliton trapped inside the structure.

\begin{figure}[b!]
\kern2\smallskipamount
\centerline{\kern-0.2em\includegraphics[width=0.2285\textwidth]{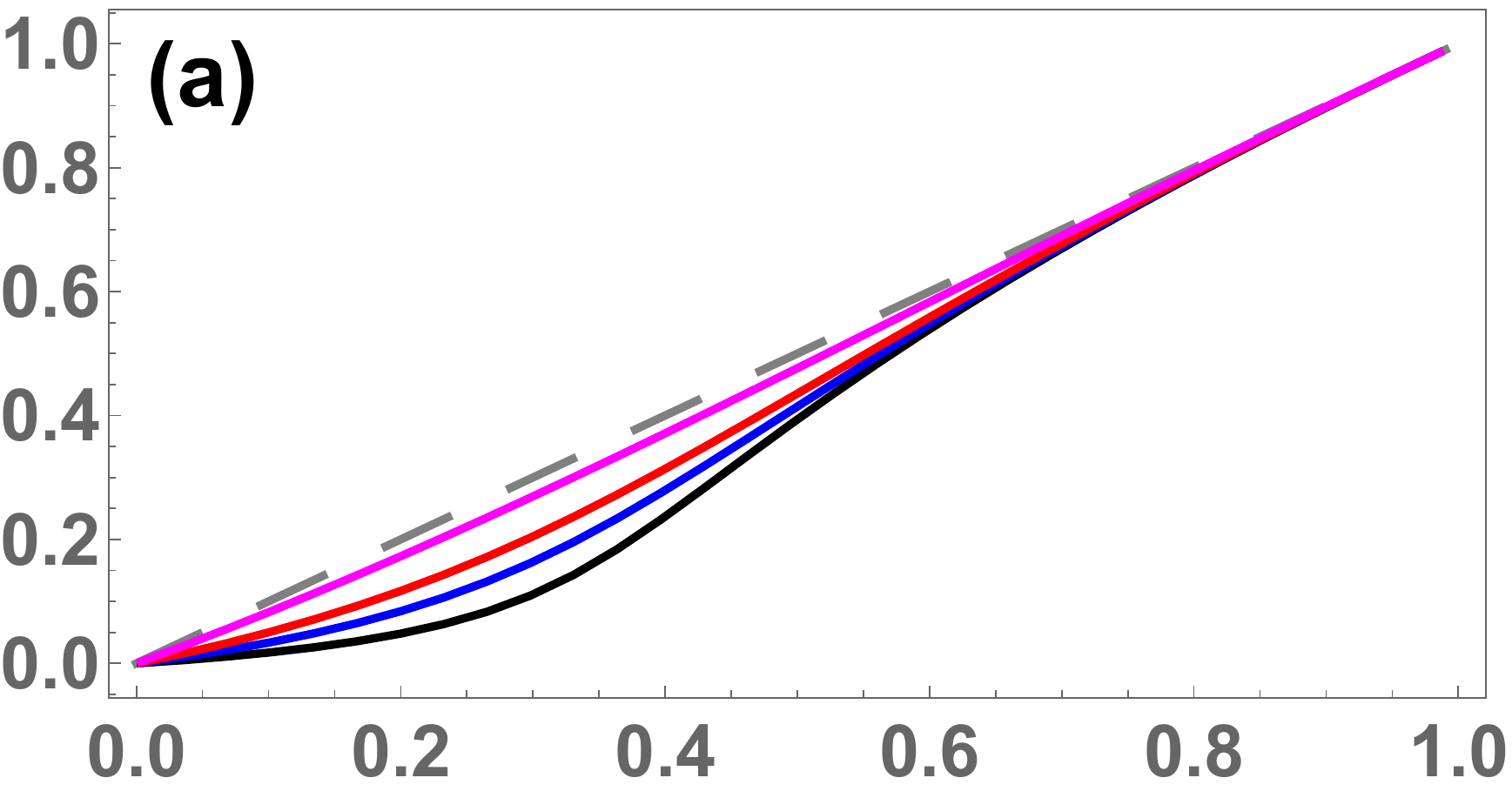}~~
\includegraphics[width=0.2285\textwidth]{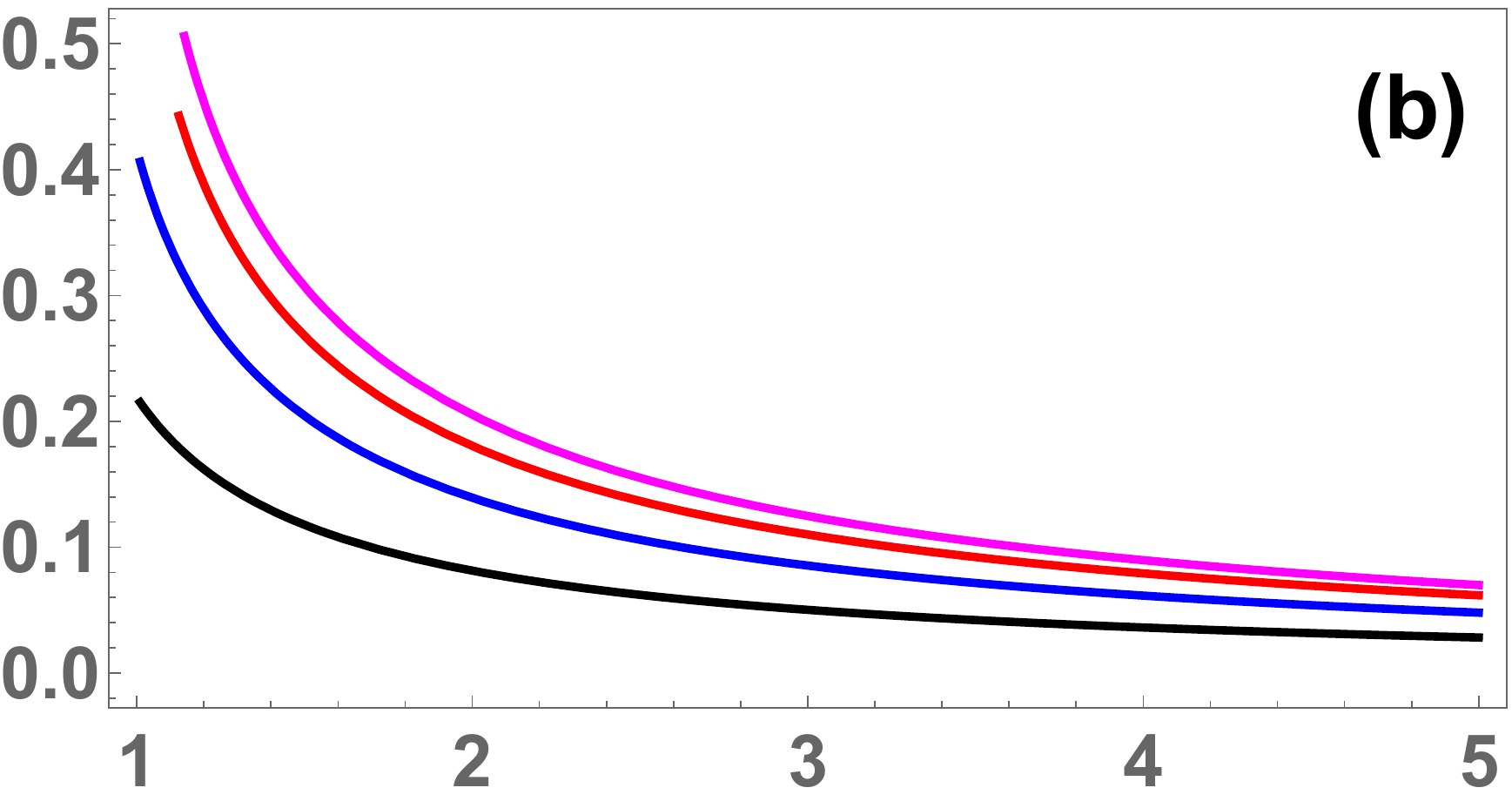}}
\caption{%
Effects of the interaction between the soliton and the wedge.
\beginblue(a) \endblue Velocity $V_*$ (vertical axis) of the trapped soliton after it enters the wedge as a function of the incident soliton velocity $V_o$.
(Black: $k_\im = 1.02$; blue: $k_\im = 1.05$; red: $k_\im = 1.1$; magenta: $k_\im = 1.4$.
Also shown for comparison is the dashed gray line $V_* = V_o$.)
\beginblue(b) \endblue Velocity $V_*$ of the soliton-generated wake as a function of $V_o$.
(Black: $k_\im = 0.8$; blue: $k_\im = 0.6$; red: $k_\im = 0.4$; magenta: $k_\im = 0.2$.)
All velocities are in units of $4\sqrt2q_o$.
}
\label{f:4}
\kern-2\smallskipamount
\end{figure}

At the same time, Fig.~\ref{f:3} also clearly shows that the soliton velocity inside the wedge differs from $V_o$.
To understand this phenomenon, 
recall that the calculation of the \beginblue long-time asymptotics \endblue changes for $0<\xi<4\sqrt2q_o$ \cite{biondinimantzavinos,CPAM2017}.
Specifically, in that range
the controlling phase in the \beginblue Riemann-Hilbert problem \endblue must be modified in order to regularize the problem, 
and one must replace $\Theta(k,\xi)$ with a new phase function $h(k,\xi)$ defined in terms of Abelian integrals \cite{CPAM2017} (see Appendix for details).
It is thus $h(k,\xi)$, not $\Theta(k,\xi)$,
that controls the soliton velocity $V_*$ inside the wedge.
More precisely, setting $h_\im(k_o,\xi) = 0$ yields an implicit equation that determines $V_*$ (see Appendix for details).
As shown in Fig.~\ref{f:3}\beginblue(a)\endblue, the value of $V_*$ predicted by the \beginblue long-time asymptotics \endblue 
is in excellent agreement with the numerical results.
Figure~\ref{f:4}\beginblue(a) \endblue shows a plot of $V_*$ as a function of $V_o$.
Note that $V_*$ is always less than $V_o$, meaning that the soliton always slows down as a result of the trapping.
The change in the soliton velocity can be interpreted as the result of the interaction between the soliton 
and the infinite number of sech-like peaks inside the wedge \cite{biondinilimantzavinos}.
At $\xi = V_*$, the solution is locally a nonlinear superposition of a soliton and a periodic solution of the NLS equation,
similarly to \cite{trogdondeconinck}.
Note that $V_*$ does not depend just on $V_o$, as demonstrated by Fig.~\ref{f:4}, which shows
solitons with the same $V_o$ but generated by different eigenvalues yield different values of $V_*$ in general.

\paragraph{Mixed regimes and soliton wake.}
Yet a different scenario arises when $k_o\in D_3$ 
\blue{or $k_o\in D_2^-$.
When $k_o\in D_3$,}
we again have $V_o>4\sqrt2q_o$, implying $\xi = V_o$ occurs in the plane wave region, so again we expect a soliton transmission
with an unchanged soliton velocity after the interaction.
The twist, however, is that when $k_o\in D_3$ the implicit equation $h_\im(k_o,\xi)=0$ 
also has a solution for \blue{$\xi = \xi_*\in(0,4\sqrt2q_o)$.}
Thus, the \beginblue long-time asymptotics \endblue predicts that 
\textit{a single discrete eigenvalue $k_o\in D_3$ 
generates two distinct contributions to the long-time behavior, propagating at different velocities: 
\blue{$\xi_*$} and $V_o$.}
We are unaware of any previous instances in which a similar phenomenon was reported in any physical context.

The \beginblue above \endblue predictions are borne out by the numerical simulations, 
as shown in Fig.~\ref{f:2}\beginblue(a) \endblue and Fig.~\ref{f:3}\beginblue(b)\endblue.
Both figures clearly show that the soliton is transmitted through the wedge, but at the same time a soliton-generated wake 
is clearly visible inside the wedge.
Also, Fig.~\ref{f:3}\beginblue(b) \endblue
demonstrates once more an excellent agreement with the prediction of the velocity of the soliton-generated wake coming from the \beginblue long-time asymptotics\endblue.
A plot of the velocity $V_*$ of the soliton-generated wake as a function of $V_o$
is shown in Fig.~\ref{f:4}\beginblue(b)\endblue.
Physically, the difference between $D_1$ and in $D_3$ is that,
for the same amplitude and velocity, eigenvalues in $D_3$ always lead to much broader solitons than eigenvalues in $D_1$
(cf.\ Appendix).

\blue{A similar situation arises when $k_o\in D_2^-$, 
except that no soliton arises outside the wedge in the \beginblue long-time asymptotics\endblue, 
and two contributions are generated inside the wedge.  
That is, $h_\im(k_o,\xi)=0$ twice for $\xi\in(0,4\sqrt2q_o)$: 
once for $\xi = V_*$, corresponding to the trapped soliton, 
and again for $\xi = \xi_*$, corresponding to the soliton-generated wake
(cf.\ Appendix).}

\paragraph{Discussion.}
\blue{In summary, we presented a study of nonlinear interactions between solitons and localized disturbances in focusing media with 
nonzero background, we classified the possible outcomes, 
which include soliton transmission, soliton trapping and the generation of a soliton wake,
and we identified precise conditions that determine the velocity of the trapped solitons and the soliton wake.}

We emphasize that 
in the pure transmission regime, the interaction is ``clean''.
That is, no permanent residuals are left in the wedge,
and \blue{(apart from a constant phase shift)}
the solution for $|x|<4\sqrt2q_ot$ is virtually undistinguishable from one without the soliton
(as confirmed by the analogue of Fig.~\ref{f:3} for a soliton transmission~\cite{supplement}).
The same is true in the trapping regime except for the presence of the soliton inside the structure (cf.\ Fig.~\ref{f:1}).
Conversely, in the mixed regimes the wedge acquires additional permanent localized traveling objects,
and interacts nonlinearly with them. 
The \beginblue analytical \endblue predictions for the velocity of the soliton-generated wake remain valid in the limit 
when the discrete eigenvalue $k_o\in D_3$ touches the segment $[0,iq_o]$,
in which case the soliton is replaced by an Akhmediev breather and $V_*=0$ \cite{supplement}.

Soliton interactions \blue{with NZBG} in the absence of localized disturbances have recently been studied in \cite{gelash,LiBiondini}.
Interactions between solitons and localized disturbances had been previously studied in the case of zero background 
\cite{AS1981,NMPZ1984,gordon,kuznetsovmikhailov},
but in that case they only lead to small effects, even in modulationally unstable media.  
It is only in focusing media with NZBG that dramatic effects such as \blue{a soliton velocity change
and a soliton-generated wake} arise.

The phenomena discussed here are related to interactions between solitons and
dispersive shock waves (DSWs) \cite{elhoefer}.
Soliton tunneling was recently studied in the context of DSWs in defocusing media in \cite{sprenger}.
Also, soliton trapping by an initial discontinuity was recently studied in \cite{maiden,ablowitzluocole}.
Note, however, that phenomena governed by the Korteweg-deVries (KdV) equation or the defocusing NLS equation 
are very different from those described by the focusing NLS equation.
In both the KdV equation and the defocusing NLS equation, the soliton velocity is proportional to its amplitude.
Thus, the soliton velocity and amplitude decrease as the soliton travels up the ramp generated by the discontinuity,
with the soliton eventually disappearing altogether.
In the focusing NLS equation, in contrast, the soliton velocity and the soliton amplitude are entirely decoupled 
(i.e., independent of each other), the mechanism which drives the change in the soliton velocity is completely different,
and the soliton amplitude and velocity inside the wedge are constant.

We also emphasize that the applicability of our results is expected to be very broad,
since, similarly to \cite{biondinimantzavinos,CPAM2017,biondinilimantzavinos,SIREV}, 
the results are essentially independent of the specific details of the initial localized disturbance.
Moreover, since the NLS equation arises in many physical contexts, including  
nonlinear optics, deep water waves, acoustics, plasmas and Bose-Einstein condensates,
the results of this work apply to all of these areas.
\blue{(Recall that suitable scalings to observe NLS dynamics in each of these domains
are well known, e.g., see \cite{DudleyTaylor,AS1981,Agrawal2007,IR2000,PS2003}.)
Finally, since the results of \cite{biondinimantzavinos} were shown in \cite{SIREV} to extend to a broader class of 
modulationally unstable systems, we expect that the same will apply in this case.}
In particular, nonlinear optical fibers and gravity waves in one-dimensional deep water channels
are especially promising candidates for the experimental verification of the phenomena described here.

\paragraph{Acknowledgments.}
We thank M. J. Ablowitz, M. A. Hoefer, M. Onorato and S. Trillo for many interesting discussions.
This work was partially supported by the National Science Foundation under grant numbers DMS-1614623 and DMS-1615524.

\kern-\medskipamount
\beginblue
\section*{Appendix}

Here we give a few details of the inverse scattering transform (IST) for the focusing NLS equation with NZBG, the wedge structure, the calculation of the long-time asymptotics, 
the numerical methods used and some further numerical results.

\paragraph{IST with NZBG: Direct problem.}
The direct problem in the IST consists in computing the scattering data (i.e., reflection coefficient, discrete eigenvalues and norming constants)
from the potential $q(x,t)$.
This is done through the Jost eigenfunctions $\phi_\pm(x,t,k)$, 
which are the simultaneous matrix solutions $\phi_\pm(x,t,k)$ of both parts of the Lax pair that reduce to plane waves as $x\to\pm\infty$.
In the case of NZBG, \cite{biondinikovacic}, they are given by Eq.~\eqref{e:jost},
where $\pm i\lambda$ and $E_\pm(k)$ 
are respectively the eigenvalues and corresponding eigenvector matrices of $X_\pm = \lim_{x\to\pm\infty}X$.
The value of $\lambda(k)$ is uniquely determined for all $k\in\Complex$ by requiring that
the branch cut is on $i[-q_o,q_o]$, $\lambda(k)$ is continuous from the right on the cut,
and that $\Im\lambda(k)>0$ for $\Im k>0$.
These Jost eigenfunctions, which are the nonlinearization of the Fourier modes, 
are defined for all values of $k\in\Complex$ such that $\lambda(k)\in\Real$, 
namely the continuous spectrum $k\in\Sigma = \Real\cup i[-q_o,q_o]$.
The reflection and transmission coefficients are $r(k) = -a_{21}/a_{11}$
and $\tau(k) = 1/a_{11}$, respectively.
The zeros of $a_{11}(k)$ and $a_{22}(k)$ define the discrete spectrum of the problem, which leads to solitons.
As usual, the time evolution in the IST is trivial.  In particular, 
with the above normalization of the Jost eigenfunctions, all the scattering data are independent of time \cite{biondinikovacic}.

\paragraph{IST with NZBG: Inverse problem.}
The inverse problem in the IST consists in reconstructing the solution $q(x,t)$ of the NLS equation from the scattering data, 
and is formulated in terms of a Riemann-Hilbert problem, 
namely the problem of finding the sectionally meromorphic matrix  
$M(x,t,k)$ which in terms of the direct problem is given by \cite{biondinimantzavinos,CPAM2017}
\vspace*{-1ex}
\begin{gather}
M(x,t,k) = 
\begin{cases}
\bigg(\displaystyle\frac{\phi_{+,1}}{a_{22}d}\,,\,\phi_{-,2}\bigg)\,\e^{-i\theta\sigma_3},~~ k\in\Complex^+\setminus i[0,q_o],
\\
\bigg(\displaystyle\phi_{-,1}\,,\,\frac{\phi_{+,2}}{a_{11}d}\bigg)\,\e^{-i\theta\sigma_3},~~ k\in\Complex^-\setminus i[-q_o,0],
\end{cases}
\end{gather}
where
$\Complex^\pm$ 
are the upper half and lower half of the complex $k$-plane, respectively, 
$\phi_{\pm,j}$ for $j=1,2$ denote the columns of $\phi_\pm$
and 
$d(k) = 2\lambda/(k+\lambda) = \det E_\pm(k)$.
More precisely, the RHP consists in computing $M(x,t,k)$ 
from the knowledge of the jump condition,
\be
M^+(x,t,k) = M^-(x,t,k)V(x,t,k),\quad k\in\Sigma,  
\ee
where superscripts $\pm$ denote projection from the left/right of the contour $\Sigma$
(oriented rightward along the real $k$-axis and upward along the segment $i[-q_o,q_o]$),
together with the normalization
$M(x,t,k) = I +O(1/k)$ as $k\to\infty$,
residue conditions at the discrete eigenvalues
and suitable growth conditions at the branch points
\cite{BilmanMiller}.
Note that $\det M(x,t,k) = 1$ for $k\in\Complex\setminus\Sigma$.
The jump matrix, obtained using the scattering relation and symmetries, is \cite{CPAM2017}
\vspace*{-0.4ex}
\bse
\begin{gather}
V(x,t,k) = \begin{pmatrix}g/d &\=r\,\e^{2i\theta} \\ r\,\e^{-2i\theta} & d 
  \end{pmatrix},
\quad k\in\Real,
\\
V(x,t,k) = \begin{pmatrix} i(\lambda-k)\=r/q_-\e^{2i\theta} & - 2i\lambda/q_-^* \\ 
  - iq_-^*g/(2\lambda) &  i(\lambda+k)r/q_-^*\,e^{-2i\theta} 
  \end{pmatrix}\!,
\nonumber\\
\kern18.2em
\quad k\in i[0,q_o],
\\
V(x,t,k) = - \begin{pmatrix} i(\lambda+k)\=r/q_-\e^{2i\theta} & iq_-g/(2\lambda) \\
 2i\lambda/q_- & i(\lambda-k)r/q_-^*\e^{-2i\theta}  
  \end{pmatrix},
\nonumber\\
\kern18em
\quad k\in i[-q_o,0],
\end{gather}
\ese
with $\=r(k) = r^*(k^*)$ and $g(k) = 1 + r(k)\=r(k)$,
where the asterisk denotes complex conjugation.
The residue condition at the poles induced by the discrete eigenvalues are \cite{biondinikovacic}
\vspace*{-1ex}
\bse
\begin{gather}
\Res\limits_{k=k_o}M(x,t,k) = M(x,t,k_o) \begin{pmatrix} 0 & 0 \\ c_o &0 \end{pmatrix}\e^{-2i\theta(x,t,k_o)},
\\
\Res\limits_{k=k_o^*}M(x,t,k) = M(x,t,k_o^*) \begin{pmatrix} 0 & -c_o^* \\ 0 &0 \end{pmatrix}\e^{-2i\theta(x,t,k_o^*)},
\end{gather}
\ese
$c_o$ being an arbitrary, complex-valued norming constant.

As shown in \cite{biondinifagerstrom}, 
the signature of MI in the inverse problem is the exponentially growing entries of $V(x,t,k)$ 
for $k\in i[-q_o,q_o]$
through the time dependence of $\theta(x,t,k)$.

\paragraph{The coherent oscillation region.}
As shown in \cite{biondinilimantzavinos,CPAM2017}, when no solitons are present
the leading-order solution in the coherent oscillation region is expressed in terms of Jacobi elliptic functions,
and represents a slow modulation of the traveling wave (periodic) solutions
of the focusing NLS equation.
In particular,
\vspace*{-1ex}
\begin{multline}
|q_{\mathrm{asymp}}(x,t)|^2 = (q_o + \alpha_\im)^2
\\
- 4q_o\alpha_\im\,\mathop{\rm sn}\nolimits^2[C_+(x-2\alpha_\re t-X);m],
\end{multline}
where the elliptic parameter $m$ and the constants $C_\pm$ are 
\vspace*{-0.6ex}
\bse
\begin{gather}
m = 4q_o\alpha_\im/[\alpha_\re^2 + (q_o + \alpha_\im)^2],
\label{e:mdef}
\\
C_\pm = |\alpha \pm iq_o| = \sqrt{\alpha_\re^2 + (q_o \pm \alpha_\im)^2}, 
\end{gather}
\ese
and the slowly varying offset $X$ is explicitly determined by the reflection coefficient. 
The four points $\pm iq_o$, $\alpha = \alpha_\re + i\alpha_\im$ and $\alpha^*$ are the branch points of
the elliptic solutions of the focusing NLS equation.
In particular, $\alpha$ is a slowly varying function of~$\xi$
determined implicitly via the system of modulation equations \cite{el,kamchatnov}
\vspace*{-0.6ex}
\bse
\label{e:modulationsystem}
\begin{gather}
\xi = 4\alpha_\re + 2(q_o^2-\alpha_\im^2)/{\alpha_\re}\,,
\\[0.4ex]
\big(\alpha_\re^2 + (q_o-\alpha_\im)^2\big)K(m) = (\alpha_\re^2-\alpha_\im^2+q_o^2)E(m)\,,
\label{e:modulation2}
\end{gather}
\ese
where $K(m)$ and $E(m)$ are the complete elliptic integrals of the first and second kind
\cite{NIST},
respectively.
%
%
The detailed properties of the asymptotic state in the coherent oscillation region
were characterized in~\cite{biondinilimantzavinos}.
\blue{
When $\xi=4\sqrt2 q_o$, Eqs.~\eqref{e:modulationsystem} yield $m=0$ and $\alpha= q_o/\sqrt2$;
when $\xi=0$, one has $m=1$ and $\alpha = iq_o$.
The trajectory described by $\alpha$ in the complex $k$-plane for $\xi\in(0,4\sqrt2q_o)$ is given by the red curve in Fig.~\ref{f:2} in the main text,
which delimits the upper and lower portions of $D_2$,
respectively denoted $D_2^\pm$.}

\paragraph{Long-time asymptotics with a discrete spectrum.}
%
%
Consider the situation in which a discrete eigenvalue at $k=k_o$ is present in the spectrum.
Owing to the invariance of the NLS equation under spatial reflections, it is enough to consider the case in which 
$k_o$ is in the first quadrant of the spectral plane,
implying $V_o>0$.
Note that the asymptotic results are independent of whether the sign of the initial soliton position $X_o$ is positive or negative,
the only difference being whether the interaction between the soliton and the localized disturbance occurs at positive versus negative times.

Let $x=\xi t$ and $\theta(x,t,k) = \Theta(k,\xi)t$, with $\Theta(k,\xi) = \lambda(\xi-2k)$.
As in \cite{CPAM2017}, the calculation of the long-time asymptotics differs depending on whether $|\xi|\gl4\sqrt2q_o$.
When $|\xi|>4\sqrt2q_o$, one can deform the RHP to remove the exponentially growing jumps
without introducing additional branch cuts or modifying the phase function. 
Conversely, when $|\xi|<4\sqrt2q_o$, in order to remove the exponential growth 
one must introduce an additional branch cut 
and replace $\Theta(k,\xi)$ with a new phase function \blue{$h(k,\xi)$} 
defined by the Abelian integral
\cite{CPAM2017}%
\vspace*{-1ex}
\bse
\label{e:hdef}
\begin{gather}
h(k,\xi) = {\frac12}\bigg(\int_{-iq_o}^k + \int_{iq_o}^k\bigg)\,\d h\,, 
\\
\d h = -4(z-\xi/4 + \alpha_\re)(z-\alpha)(z-\alpha^*)/\gamma(z)\,\d z\,
\end{gather}
\ese
with
$\gamma(z) = [(z^2+q_o^2)(z-\alpha)(z-\alpha^*)]^{1/2}$.
\blue{Without loss of generality, we take the two branch cuts of $h(k,\xi)$ respectively along the segment of the imaginary axis from $-iq_o$ to $iq_o$
and along the curve $h_\im(k,\xi)=0$ connecting $\alpha^*$ to $\alpha$ \cite{CPAM2017}.}

The only difference between the long-time asymptotics in our case and that in \cite{CPAM2017}
arises when $\Theta_\im(k_o,\xi)=0$ in the plane wave region
or $h_\im(k_o,\xi)=0$ in the wedge.
Those are the only values of $\xi$ at which the poles in the RHP give an $O(1)$ contribution to the solution.
Conversely, as in the case of zero background, for all values of $\xi$ in the plane wave region such that $\Theta_\im(k_o,\xi)\ne0$, 
and all values of $\xi$ in the wedge such that  $h_\im(k_o,\xi)\ne0$,
the poles give an exponentially small contribution to the solution as $t\to\infty$.

As discussed in the main text, 
the condition $\Theta_\im(k_o,\xi)=0$ is satisfied for $|\xi|>4\sqrt2q_o$ when the discrete eigenvalue $k_o$ 
lies in $D_1$ (transmission region) or $D_3$ (wake region), leading to the appearance of a soliton in the plane wave region 
in those cases, but not when $k_o\in D_2$ (trapping region), leading to the absence of a soliton in the plane wave region in that case.
It therefore remains to examine whether the condition $h_\im(k_o,\xi)=0$ is satisfied for $|\xi|<4\sqrt2q_o$ in each of these three cases.
We discuss this issue next.

\begin{figure}[t!]
\includegraphics[width=0.235\textwidth]{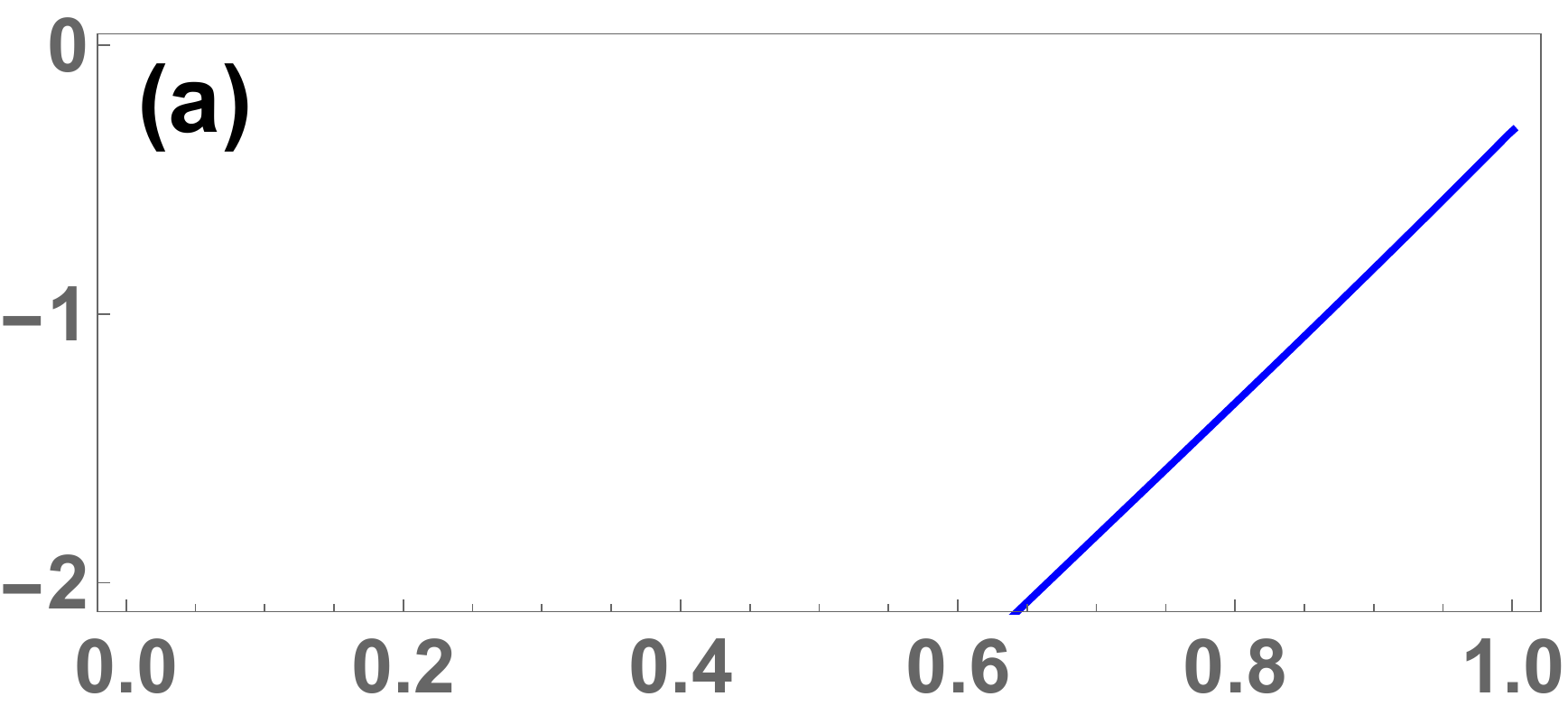}
\includegraphics[width=0.235\textwidth]{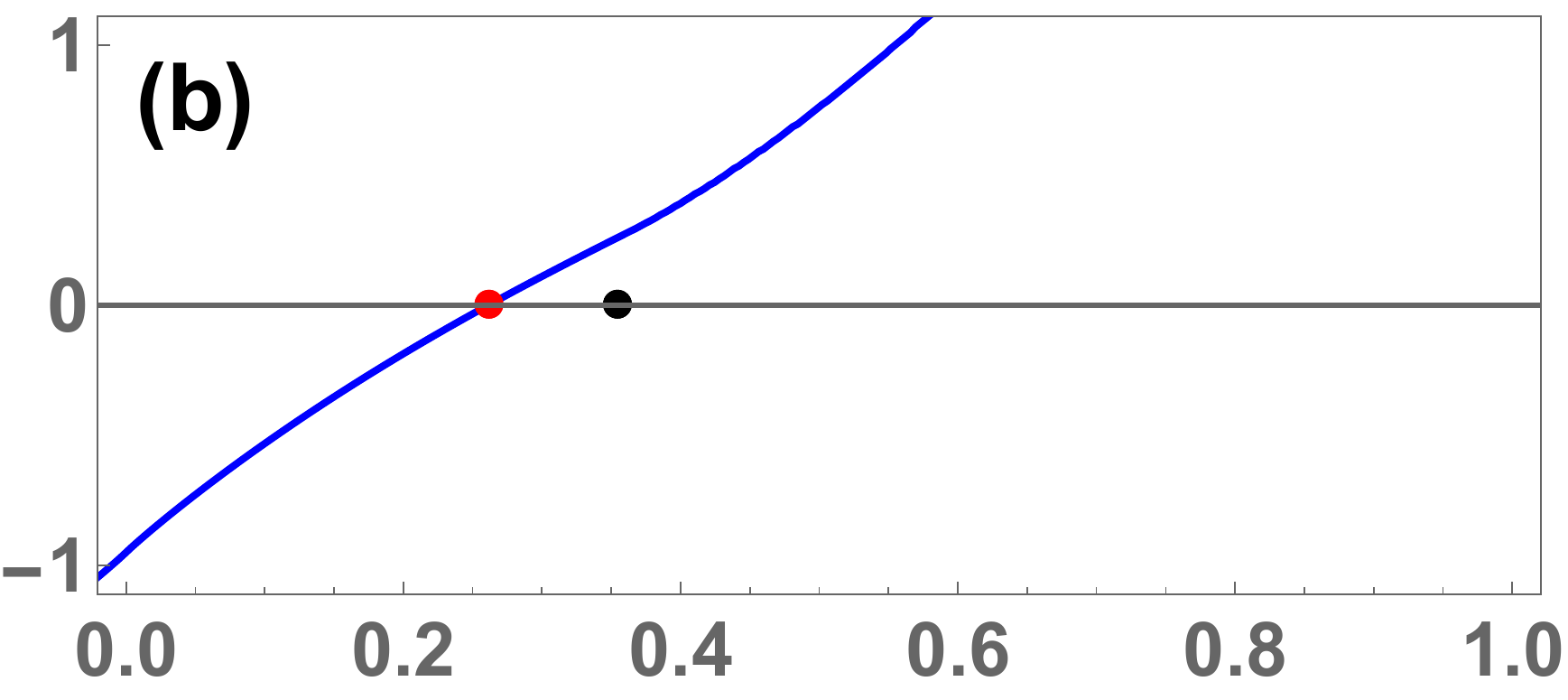}\\
\includegraphics[width=0.235\textwidth]{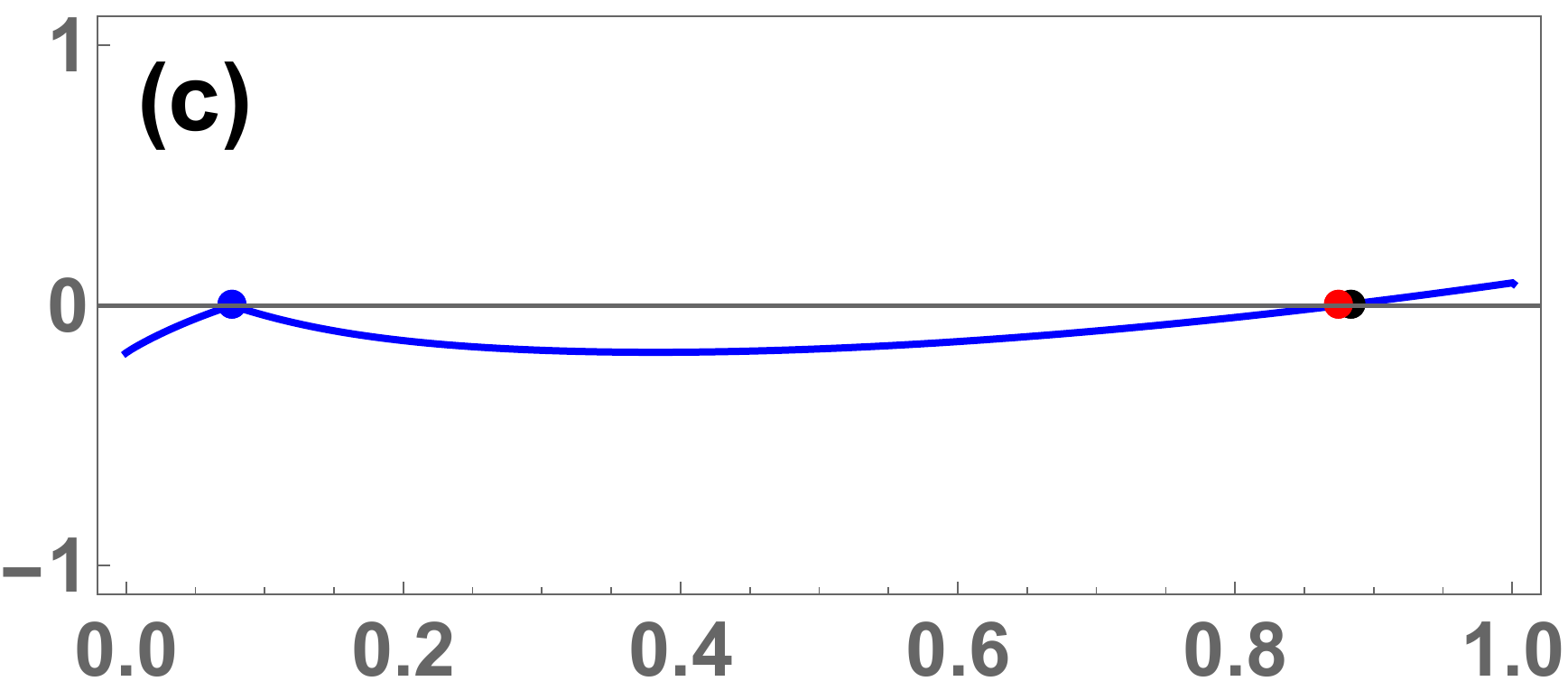}
\includegraphics[width=0.235\textwidth]{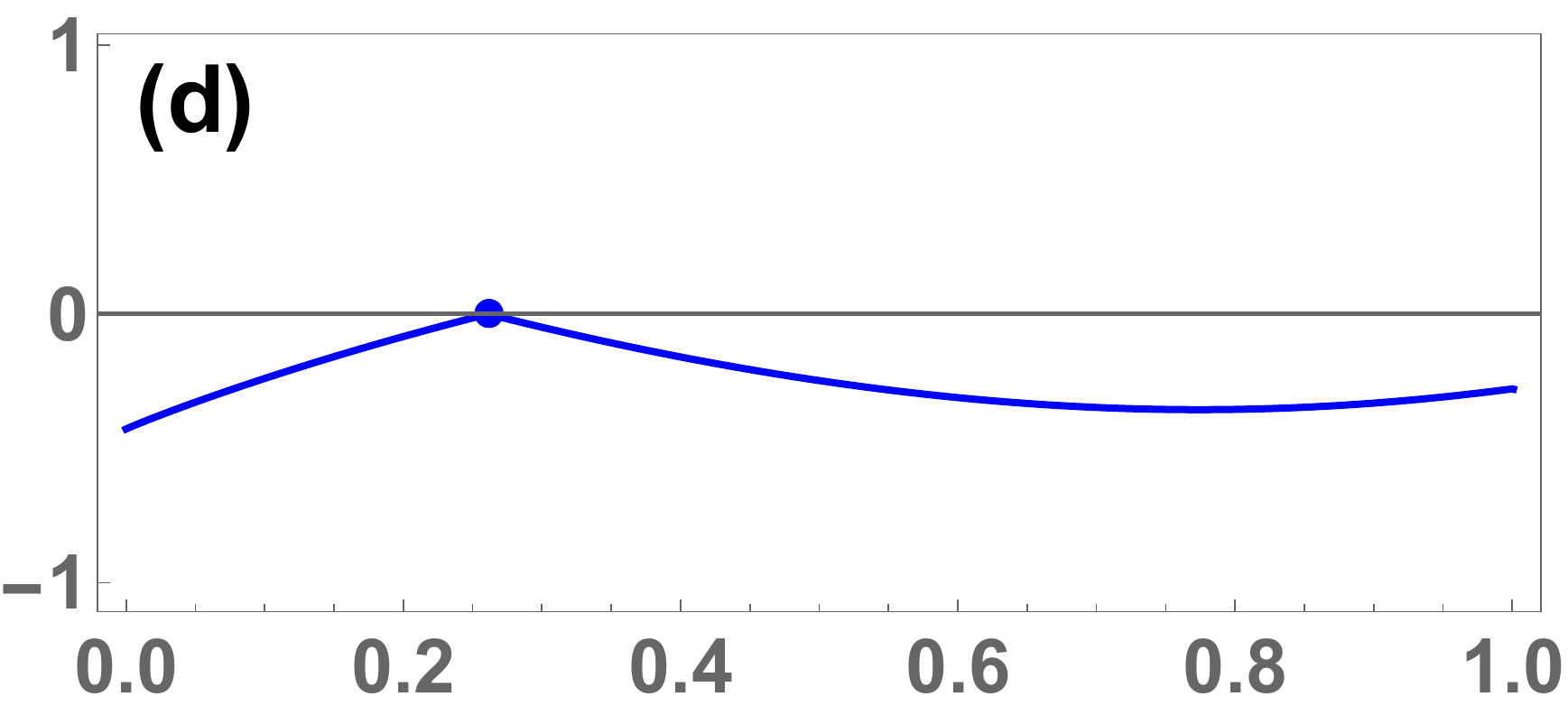}
\beginblue
\caption{The value of $h_\im(k_o,\xi)$ (blue curve) as a function of $\xi$ (in units of $4\smash{\sqrt{2}}q_o$) 
in the wedge in a few representative cases.
(a) 
$k_o = 1.24 + 1.1 i\blue{\in D_1}$ ($V_o=6$, transmission).
(b) 
$k_o = 0.216 + 1.1 i\blue{\in D_2^+}$ ($V_o=2$, \blue{trapping}).
(c) 
$k_o = \blue{0.0479 + 0.95 i\in D_2^-}$ ($V_o=\blue{5}$, \blue{trapping plus wake}).
(d) 
$k_o = 0.214 + 0.5 i\blue{\in D_3}$ ($V_o=8$, \blue{transmission plus wake}).
Black dots: $V_o$;
red dots: $V_*$ \blue{(trapped soliton); blue dots: $\xi_*$ (wake).}
}
\label{f:S1}
\endblue
\kern-\medskipamount
\end{figure}

\paragraph{Implicit equation for the soliton velocity \blue{in the wedge.}}
The integrals in Eq.~\eqref{e:hdef} can be carried out and expressed in terms of incomplete elliptic functions.
\blue{On the other hand, we found it more convenient to just evaluate numerically the imaginary part of $h$, 
which can be shown to be simply}
\blue{\[
h_\im(k,\xi) =  \frac1{2i}\int_{k^*}^k \d h\,.
\]
}
\unskip
The value of $V_*$ is \blue{then} computed numerically with standard root finding algorithms.

Figure~\ref{f:S1} shows the value of $h_\im(k_o,\xi)$ as a function of $\xi$ in the wedge 
(i.e., in the range $-4\sqrt2q_o<\xi<4\sqrt2q_o$) 
for a few representative \blue{values of $k_o$} in the transmission, trapping and mixed \blue{regimes}.
From these plots we see that \blue{the equation $h_\im(k_o,\xi)=0$}
has no solution when the discrete eigenvalue $k_o$ is in $D_1$,
i.e., in the \blue{pure} transmission regime.
Conversely, \blue{$h_\im(k_o,\xi)=0$} exactly \blue{once} when 
\blue{$k_o\in D_2^+$, corresponding to the trapped soliton,
and when $k_o\in D_3$, corresponding to the soliton-generated wake.
Finally, $h_\im(k_o,\xi)=0$ twice when $k_o\in D_2^-$, with one zero corresponding to the trapped soliton
and the other to the wake.}
Note that $V_*$ is always less than $V_o$ in absolute value, and 
$V_*\to 4\sqrt2q_o$ whenever $V_o\to 4\sqrt2_o$.

\begin{figure}[b!]
\centerline{\hglue-0.2em%
\includegraphics[height=0.245\textwidth]{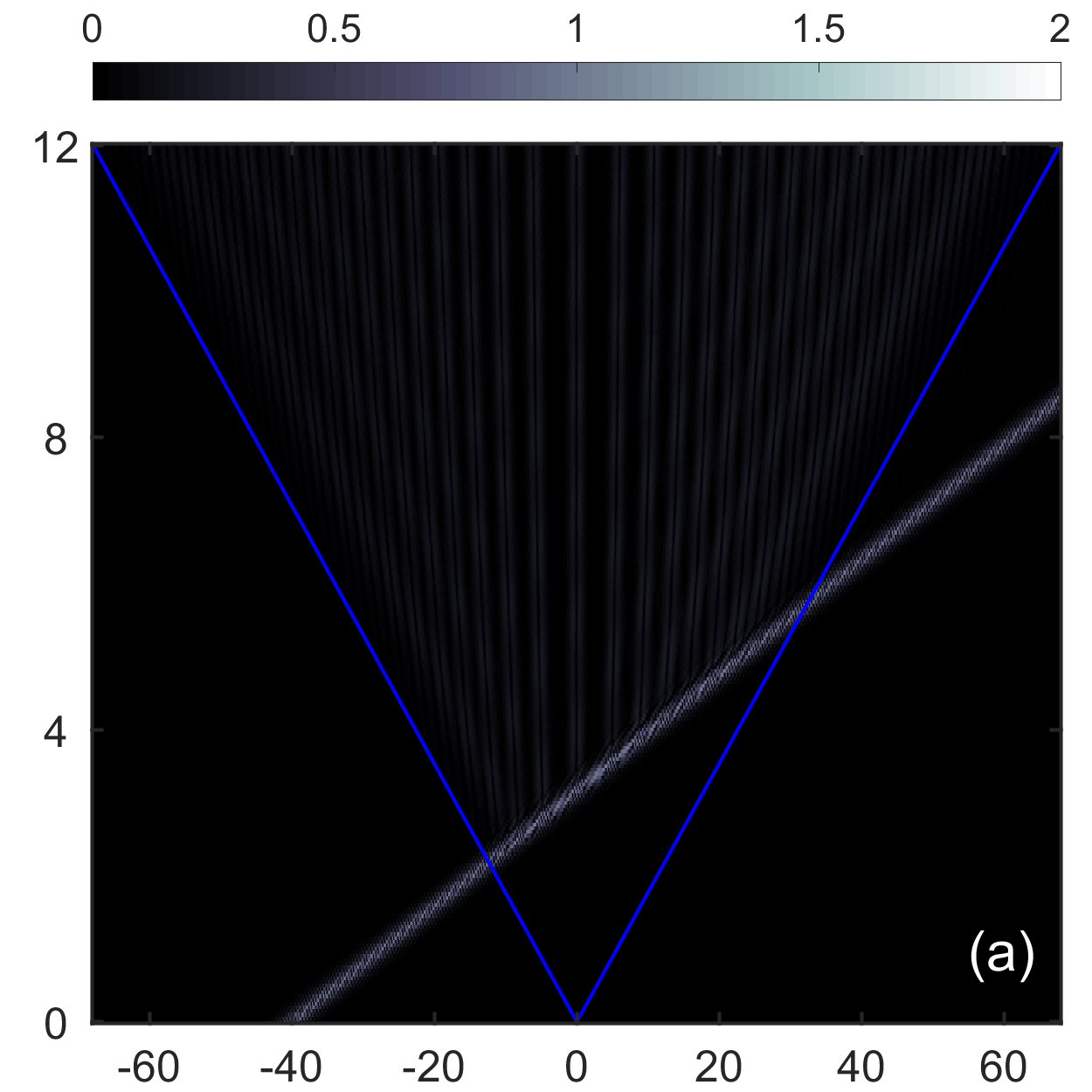}
\raise0ex\hbox{\includegraphics[height=0.23\textwidth]{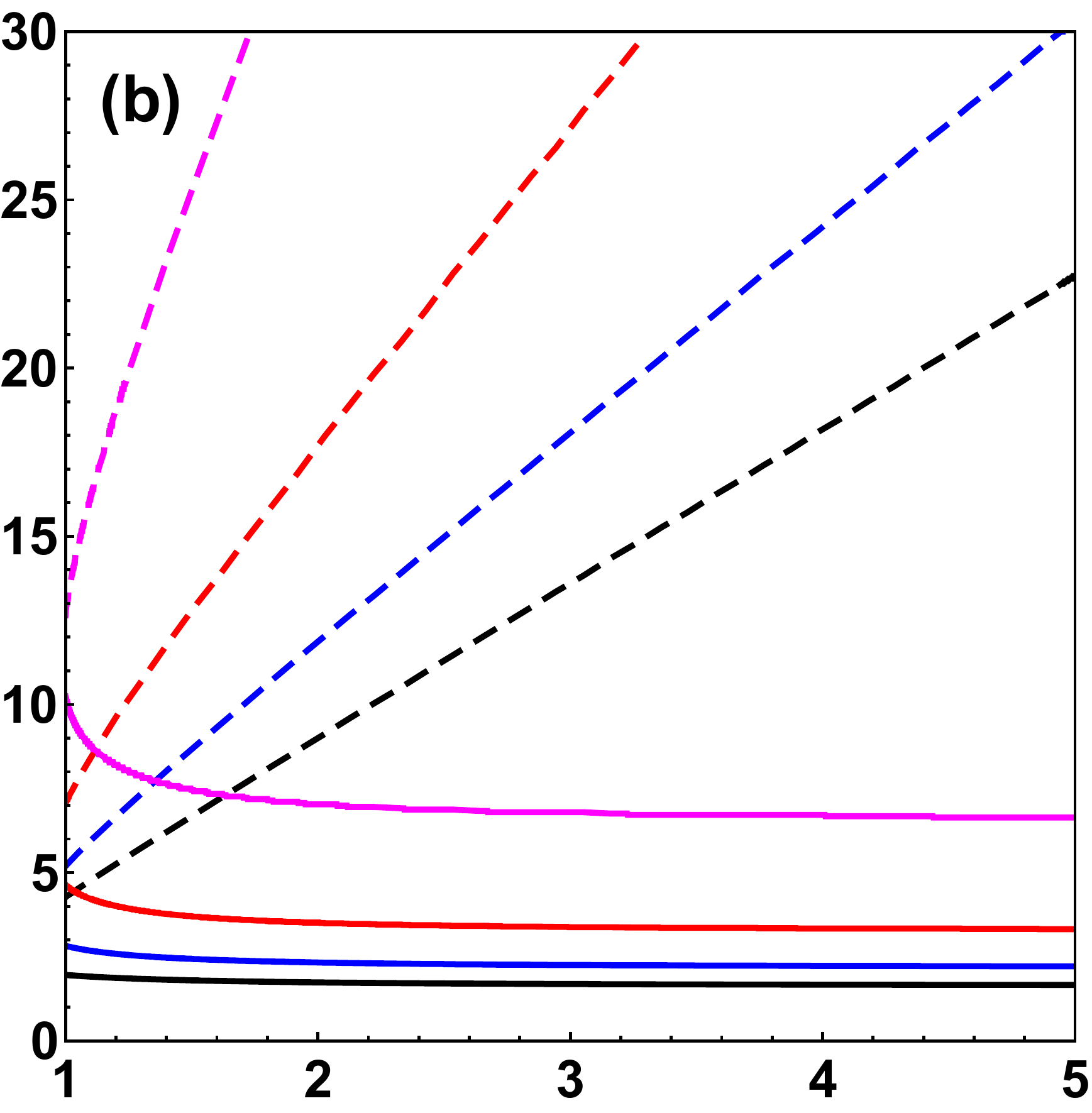}}}
%
\beginblue
\caption{(a) Amplitude difference between the solution shown in Fig.~\ref{f:1}\blue{(a)} and a solution without the corresponding soliton
(as in Fig.~\ref{f:4}), 
demonstrating that in this case the interaction produces no lasting effects on the wedge.
(b) full width at half maximum (vertical axis) of two solitons with the same velocity, 
one generated by a discrete eigenvalue $k_o\in D_1$ (solid lines) and the other by an eigenvalue $k_o\in D_3$ (dashed lines).
The horizontal axis is $V_o$ in units of $4\sqrt2q_o$.
Black lines: $k_\im = 0.8$;
Blue lines: $k_\im = 0.6$;
Red lines: $k_\im = 0.4$;
Magenta lines: $k_\im = 0.2$.
}
\label{f:S2}
\endblue
\bigskip
\centerline{\hglue-0.2em%
\includegraphics[height=0.245\textwidth]{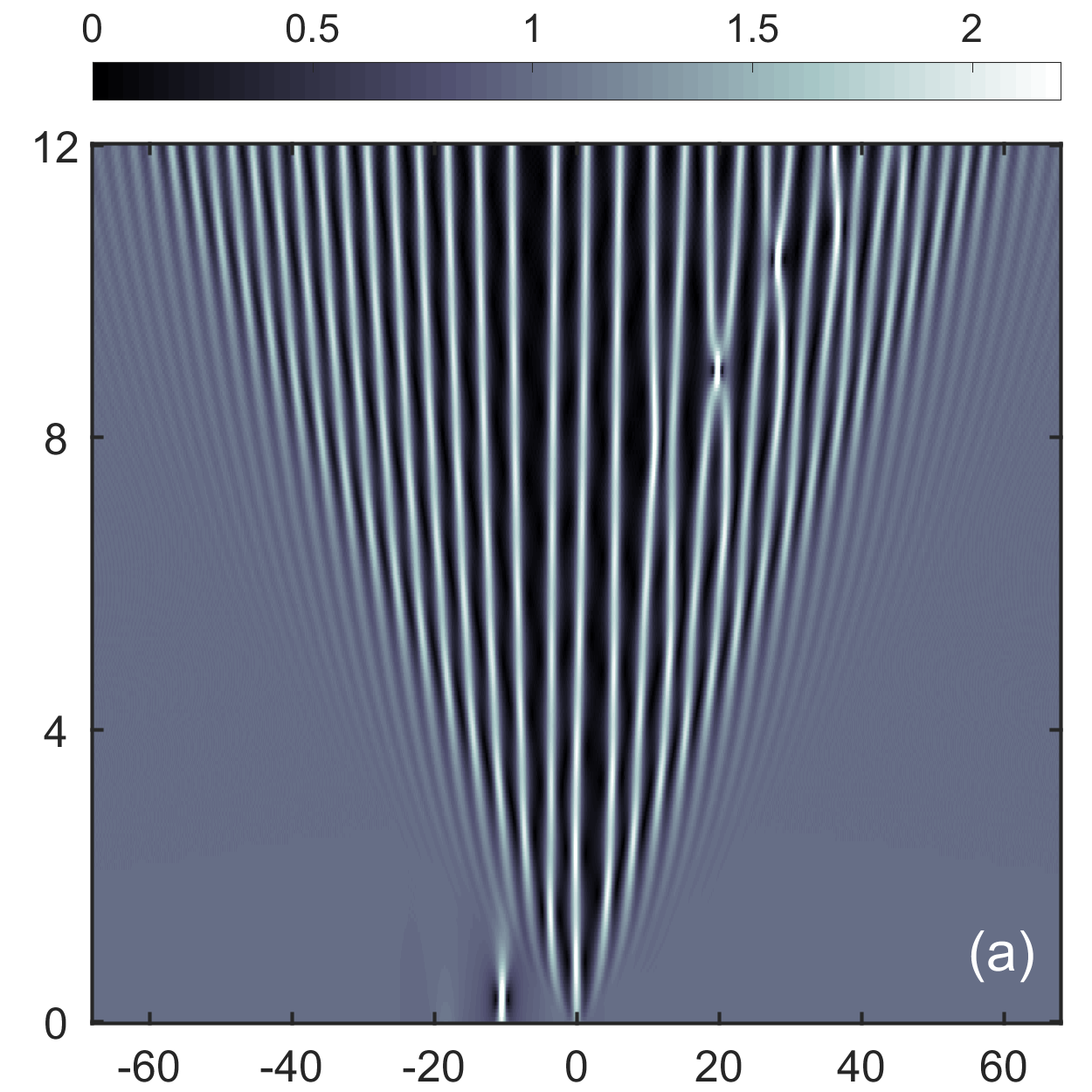}
\includegraphics[height=0.245\textwidth]{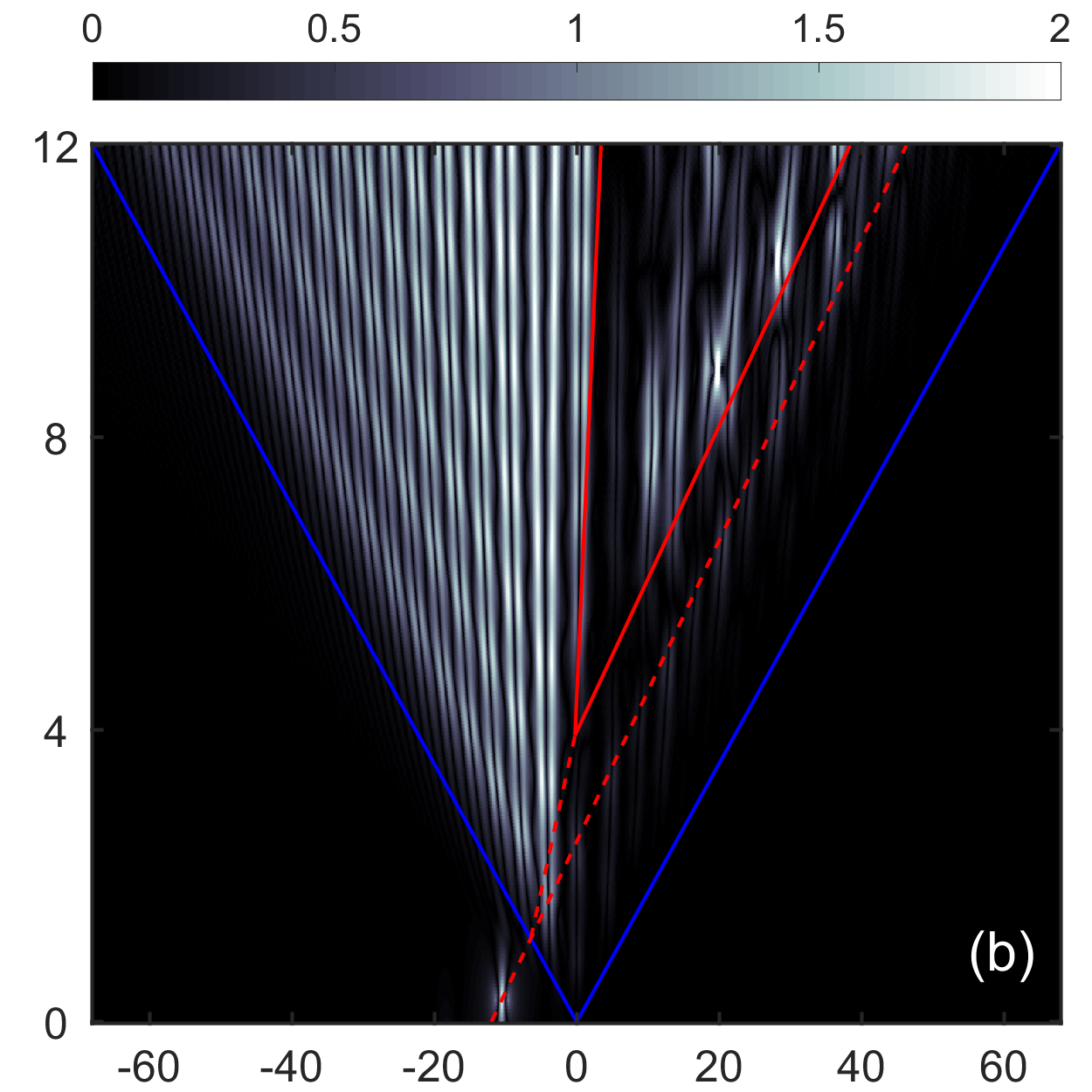}}
\beginblue
\caption{Density plot of the amplitude (a) and amplitude difference (b) produced when $k_o = 0.05+0.95i\in D_2^-$, 
resulting in a soliton trapping and a soliton-generated wake.}
\label{f:Strappingwake}%
\endblue
\bigskip
\centerline{\hglue-0.2em%
\includegraphics[height=0.245\textwidth]{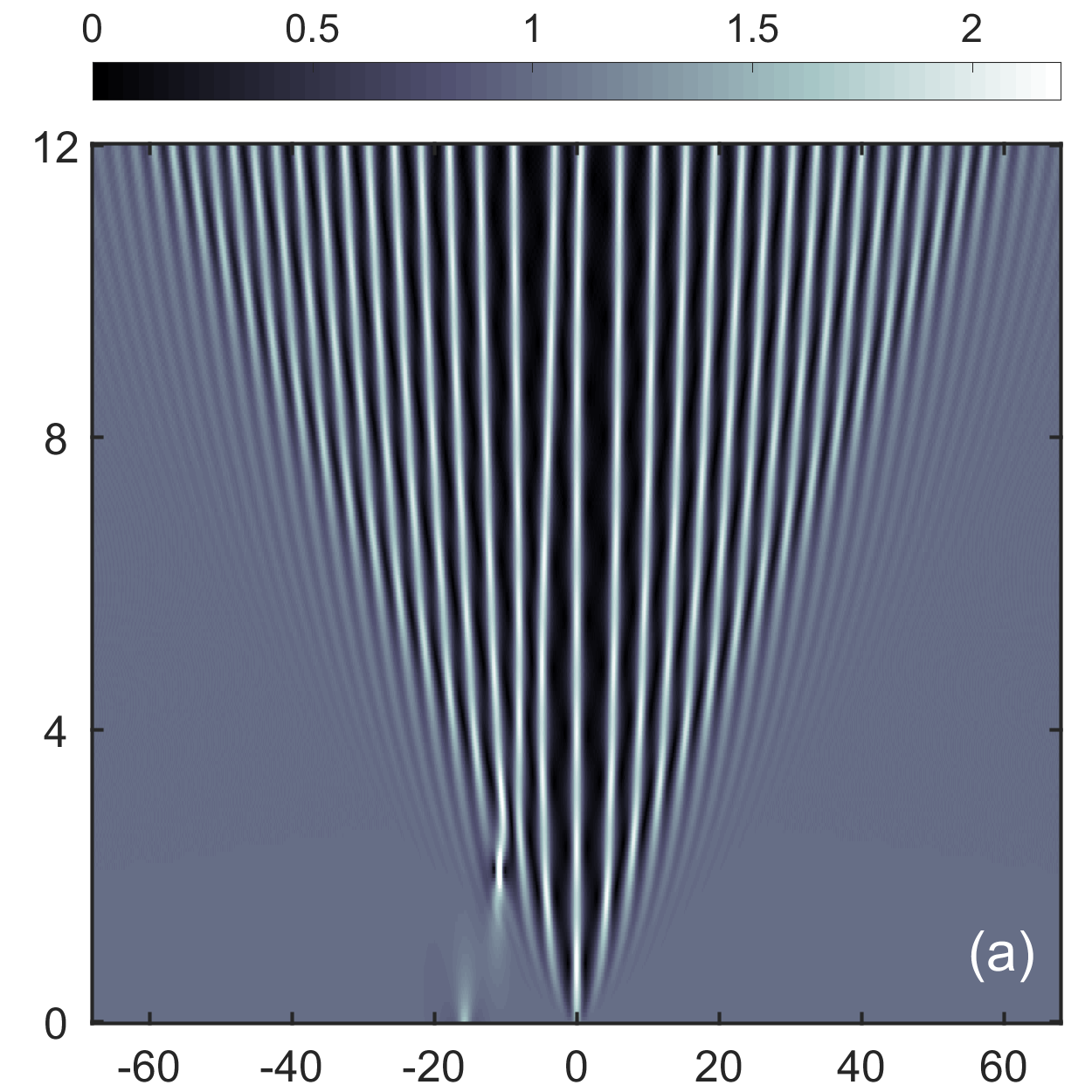}
\includegraphics[height=0.245\textwidth]{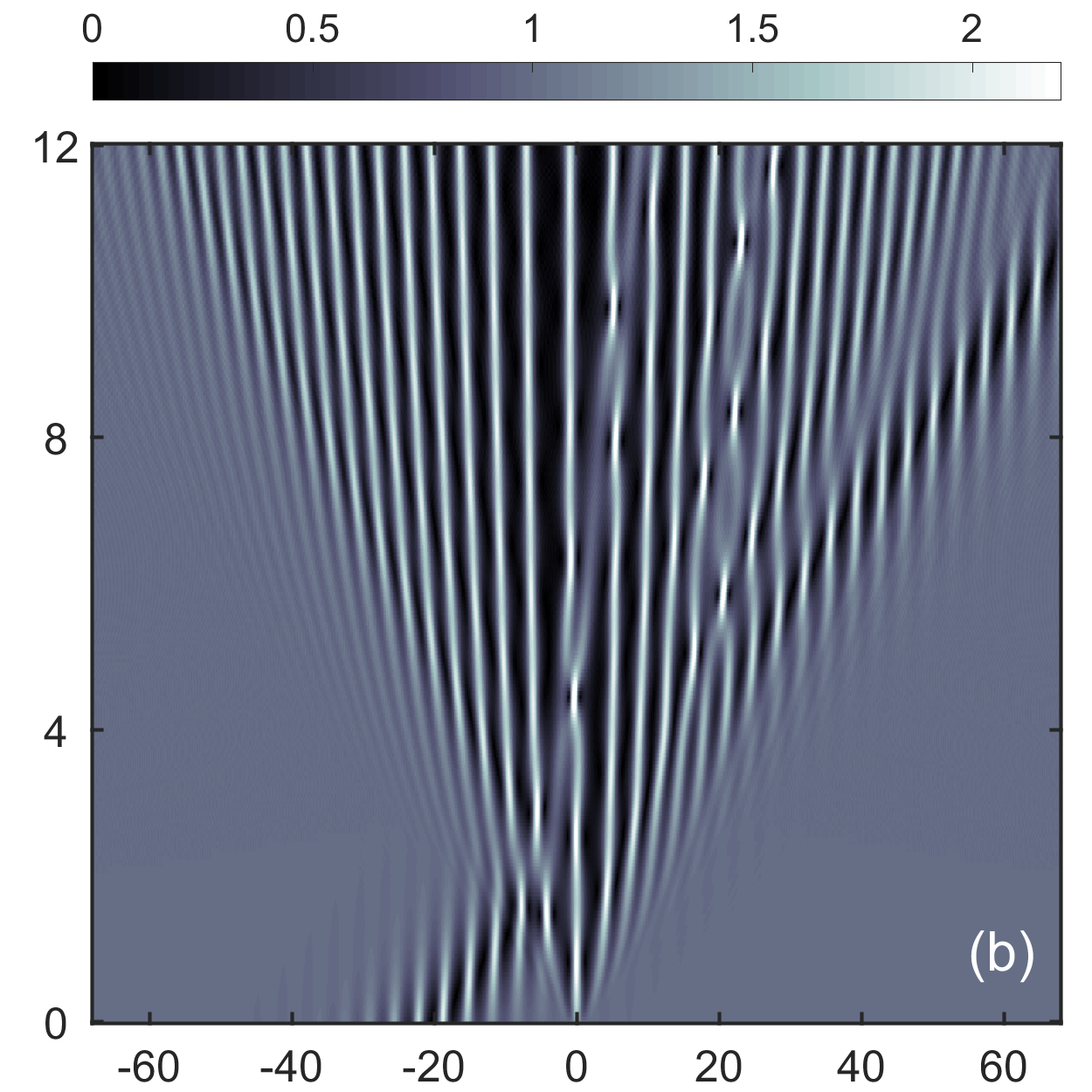}}
\beginblue
\caption{%
Density plots of the solution amplitude corresponding to Fig.~\ref{f:3} in the main text. 
}
\label{f:S3}
\endblue
\kern-1.4\medskipamount
\end{figure}

\paragraph{Numerical methods.}
All numerical simulations \blue{of the NLS equation} were performed using an eighth-order \blue{Fourier} split-step method 
\cite{yoshida}
with periodic boundary conditions
and $N=4096$ \blue{grid points}. 
The spatial domain used was much larger than the spatial window shown in the figures, 
so that the phase discontinuity at the edge of the domain generated by the soliton 
did not affect the solution in the plot window.
Specifically, we took $x\in[-L,L]$ with $L=200$, 
implying a spatial grid size of $\Delta x = 9.77\times10^{-2}$.
The initial disturbance was \blue{realized by taking}
$q(x,0) = 1 + i\e^{-x^2}\cos(\sqrt2x)$ \blue{near $x=0$.}
The time integration was performed with an integration step size of $\Delta t = 2\times10^{-4}$.
This setup allowed us to obtain accurate results until about $t=15$, at which point roundoff errors become $O(1)$.

\paragraph{Further numerical results.}
Figure~\ref{f:S2}\blue{(a)} shows the amplitude difference between the solution in Fig.~\ref{f:1}\blue{(a)} 
of the main text and a solution without the corresponding soliton,
clearly demonstrating that no permanent effects remain in the wedge.  
The right panel of Fig.~\ref{f:S2} shows a comparison between the width of two solitons with the same amplitude and velocity,
one generated by a discrete eigenvalue $k_o\in D_1$ and the other by $k_o\in D_3$, 
demonstrating that eigenvalues in $D_3$ always generate broader solitons than those in $D_1$.

\blue{For completeness, 
Fig.~\ref{f:Strappingwake} shows an interaction with a soliton generated by a discrete eigenvalue in $D_2^-$, 
in which case the interaction results in a soliton trapping and a soliton-generated wake,
and Fig.~\ref{f:S3} shows the amplitude of the solutions in Fig.~\ref{f:4} of the main text.}
Finally, we note that the long-time asymptotics results \blue{also apply} if the soliton is replaced by an Akhmediev breather, in which case \blue{one simply obtains $\xi_*=0$}.

\endblue









\input references


\end{document}